# A Metamaterial Homogenization Approach with Application to the Characterization of Microstructured Composites with Negative Parameters

Mário G. Silveirinha

Universidade de Coimbra, Instituto de Telecomunicações
Departamento de Engenharia Electrotécnica, Pólo II, 3030 Coimbra, Portugal, mario.silveirinha@co.it.pt

## Abstract

In this work, we develop a new systematic and self-consistent approach to homogenize arbitrary non-magnetic periodic metamaterials. The proposed method does not rely on the solution of an eigenvalue problem and can fully characterize the effects of frequency dispersion, magneto-electric coupling, and spatial dispersion, even in frequency band-gaps or when the materials are lossy. We formulate a homogenization problem to characterize a generic microstructured artificial material, and demonstrate that it is equivalent to an integral-differential system. We prove that this complex system can be reduced to a standard integral equation and solved using standard methods. To illustrate the application of the proposed method, we homogenize several important metamaterial configurations involving split ring resonators and metallic wires.

# I. INTRODUCTION

In the last few years there has been a growing interest in the investigation of microstructured materials, known as "metamaterials". Several interesting potentials of these artificial composites have been demonstrated/suggested, including the realization of media with negative index of refraction [1]-[2], synthesis of media that enable sub-diffraction imaging [3]-[5], the possibility of fabricating sub-diffraction cavities and waveguides [6]-[7], the possibility of squeezing electromagnetic waves through subwavelength channels [8], and many others.

What is somehow striking about the field of metamaterials is that even though some fundamental concepts such as negative permittivity/permeability are widely accepted and relatively well understood today, there is no systematic and clear way to calculate/define unambiguously these parameters for metamaterials. In many works, the calculation of $\varepsilon$ and $\mu$ is based on the classic Clausius-Mossotti (CM) homogenization formula [9]. Even though such approximation may be useful and sufficiently accurate in a number of problems, in other situations it may not be applied and completely fail. For example, one obvious problem is that the CM-formula relies on the assumption that the volume fraction of the inclusions is small, which hardly is the case for most metamaterials. Other authors rely on band structure calculations to retrieve the effective parameters of a material (e.g. [10], [11]). This procedure can be considered general and rigorous, but also suffers from important drawbacks. Namely, it is not useful to retrieve the effective parameters either in frequency band-gaps or when the constituent materials have losses. Other homogenization methods have been developed over the years [12]-[15], but in general their scope of application is limited to the static limit or to very specific geometries. The extraction of effective parameters from measured or simulated S-parameter data (e.g. [16]) is arguably the most popular method today to characterize composite media. This method has however several important drawbacks because the extraction method may be unstable, multiple solutions may occur, and most importantly, in some cases it may not be appropriate to describe thin metamaterial screens as bulk materials.

Another important aspect is that some relevant metamaterials are characterized by strong spatial dispersion [17]-[18], even for very low frequencies [19]-[20]. For example, as shown in [21], all the topologies of the 3D-wire medium are characterized by some intrinsic spatial dispersion (see also [22]). Nevertheless, to our best knowledge, there is no readily available systematic homogenization procedure that is able to predict and take into account the effect of spatial dispersion. Notice that the emergence of spatial dispersion does not preclude a material from being homogenized; it just makes things more difficult. Also we note that microstructured materials with strong spatial dispersion may have interesting applications, such as sub-diffraction imaging, as demonstrated in [4]. Another important problem is the characterization of the bianisotropic effects in metamaterials. It is known that if the basic inclusions do not have enough symmetry, then significant magneto-electric coupling (also known as optical activity [17]) may exist, being a well-known example the split-ring resonator geometry [23]. Unfortunately, there is no satisfactory theory to calculate in a systematic way the magneto-electric crossed terms and assess the effect of bianisotropy.

The main objective of this paper is to develop a new self-consistent rigorous approach to homogenize non-magnetic periodic metamaterials. The proposed method is completely general and can be used to calculate the effective parameters of arbitrary periodic dielectric/metallic metamaterials, taking into account both spatial and frequency dispersion, even in frequency bands where the propagation of electromagnetic waves is not allowed (band-gaps). The method is not based on the solution of an eigensystem and does not involve band structure calculations. Instead, we show that the homogenization problem can be formulated as a source driven problem. The idea is to excite the periodic material with a suitable source distribution. We prove that the homogenization problem is equivalent to an integral-differential system. It is shown that this system can be reduced to a standard integral equation, which can be solved with known techniques. Quite remarkably, we are able to obtain a closed-form expression for the permittivity dyadic in terms of the inverse of an infinite matrix. In order to illustrate the proposed theoretical concepts, we homogenize several relevant metamaterials with negative parameters, formed by metallic wires and split-ring resonators. In particular, we investigate the magnetic properties of closely packed arrays of split-ring resonators, and of elliptically-shaped rings, showing that the coupling between the particles may widen the bandwidth of the negative permeability region and enhance the artificial magnetism.

This paper is organized as follows. In section II, we discuss the problem of constitutive relations and the definition of effective parameters. It is argued that since spatial dispersion may not be negligible in metamaterials, these microstructured composites should preferably be characterized by a dielectric function of the form $\bar{\bar{\varepsilon}} = \bar{\bar{\varepsilon}}(\omega, \mathbf{k})$, instead of the conventional local model based on the definition of a dielectric permittivity $\varepsilon = \varepsilon(\omega)$, and permeability $\mu = \mu(\omega)$. The connections between the two models and respective scope of application are carefully discussed. In section III, we develop a microscopic theory to calculate the dielectric function $\bar{\bar{\varepsilon}} = \bar{\bar{\varepsilon}}(\omega, \mathbf{k})$. The idea is to excite the material with a source that imposes an appropriate phase modulation in the unit cell. In section IV, we use integral equation methods to solve the homogenization problem. In section V, we illustrate the application of the proposed method to the characterization of artificial materials formed by split-ring resonators and metallic wires. In particular, we study the magnetic properties of arrays of closely packed resonators and elliptically-shaped split ring resonators. It is explained how the permeability $\mu = \mu(\omega)$ can be extracted from the dielectric tensor $\bar{\bar{\varepsilon}} = \bar{\bar{\varepsilon}}(\omega, \mathbf{k})$. We compare with good agreement the homogenization results with the dispersion characteristics obtained from the calculation of the band structure of the materials. Finally, in section VI the conclusions are drawn.

In this work, we assume that the fields are monochromatic with time dependence $e^{j\omega t}$. We denote the microscopic electric and induction fields by $\mathbf{E}$ and $\mathbf{B}$. On the other hand, the macroscopic (spatially averaged) electric and induction fields are denoted by $\mathbf{E}_{av}$ and $\mathbf{B}_{av}$. Since the electric displacement vector and the magnetic field are inherently macroscopic quantities, they are denoted by $\mathbf{D}$ and $\mathbf{H}$ (the subscript "av" is omitted).

## II. CONSTITUTIVE RELATIONS

We consider a generic non-magnetic periodic metamaterial invariant to translations along the primitive vectors $\mathbf{a}_1$, $\mathbf{a}_2$, and $\mathbf{a}_3$ (see Fig. 1). The metamaterial is characterized by the (relative) permittivity $\varepsilon_r(\mathbf{r})$, where $\mathbf{r} = (x, y, z)$ is a generic point of space; the permittivity is normalized to the host permittivity $\varepsilon_h$, which for simplicity is taken to be the permittivity of vacuum $\varepsilon_0$. All the results derived in this work remain valid if $\varepsilon_0$ is replaced by $\varepsilon_h$. Since the medium is periodic the permittivity satisfies $\varepsilon_r(\mathbf{r} + \mathbf{r}_\mathbf{I}) = \varepsilon_r(\mathbf{r})$, where $\mathbf{r}_\mathbf{I} = i_1\mathbf{a}_1 + i_2\mathbf{a}_2 + i_3\mathbf{a}_3$ is a lattice point and $\mathbf{I} = (i_1, i_2, i_3)$ is a generic multi-index of integers. The unit cell $\Omega$ of the periodic medium is $\Omega = \{\alpha_1\mathbf{a}_1 + \alpha_2\mathbf{a}_2 + \alpha_3\mathbf{a}_3 : |\alpha_i| \leq 1/2\}$. The permittivity $\varepsilon_r$ may be a complex number and depend on frequency. In addition, we admit that the unit cell may contain perfectly electric conducting (PEC) metallic surfaces, which we denote by $\partial D$ (see Fig. 1). The outward unit vector normal to $\partial D$ is $\hat{\mathbf{v}}$.

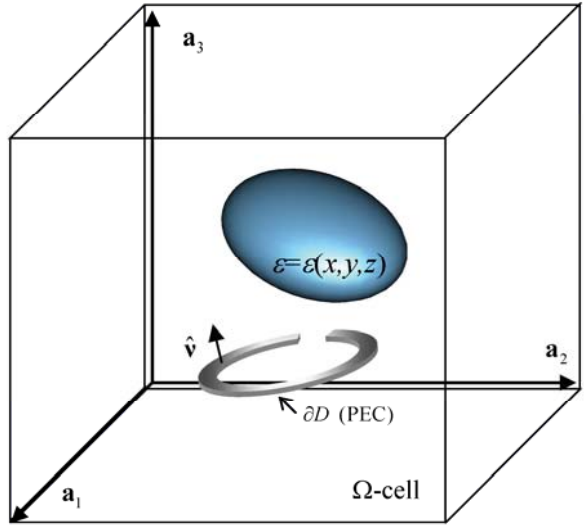

**Fig. 1** (Color online) Geometry of a generic metallic-dielectric periodic material with a dielectric inclusion and a PEC inclusion.

The objective of this work is to homogenize the periodic metamaterial and extract the effective parameters. Before developing a microscopic theory – this will be the topic of the next section – it is fundamental to carefully discuss the problem of phenomenological constitutive relations [24].

In most of the works devoted to metamaterials, the artificial structures are characterized using an effective permittivity and an effective permeability. Such description is very appealing and useful, but, as discussed next, in some cases its scope of application to metamaterials may be limited.

To understand these limitations, it is useful to briefly go over the standard derivation of macroscopic electromagnetism. This theory is based on the fact that when an external field interacts with matter it induces microscopic currents $\mathbf{J}_d$, such that its spatial average, $\langle \mathbf{J}_d \rangle$, can be expanded into dipolar and higher-order contributions [9, pp. 256]:

$$\langle \mathbf{J}_d \rangle \approx j\omega \mathbf{P} + \nabla \times \mathbf{M} + ... \qquad (1)$$

where $\mathbf{P}$ is the polarization vector and $\mathbf{M}$ is the magnetization vector. The terms that are omitted involve spatial derivatives of the quadrupole density and other higher-order multipole moments. The classical definition of the (macroscopic) electric displacement vector **D** and of the (macroscopic) magnetic field **H** is motivated by the decomposition (1) of the average microscopic current into mean and eddy currents. Indeed, **D** and **H** are derived/defined from the fundamental macroscopic fields $\mathbf{E}_{av}$ (average electric field) and $\mathbf{B}_{av}$ (average induction field) through the textbook formulas [9, pp. 248],

$$\mathbf{D} = \varepsilon_0 \mathbf{E}_{av} + \mathbf{P} \tag{2a}$$

$$\mathbf{H} = \frac{\mathbf{B}_{av}}{\mu_0} - \mathbf{M} \tag{2b}$$

Thus, definition (2) implicitly absorbs the effect of the microscopic currents into **D** and **H**, and so the macroscopic Maxwell-Equations in the material have the same form as in vacuum (apart from the relation between $\mathbf{E}_{av}$, **D**, $\mathbf{B}_{av}$, and **H**). For linear materials **P** and **M** can be written as a linear combination of $\mathbf{E}_{av}$ and **H** (or equivalently, in terms of $\mathbf{E}_{av}$ and $\mathbf{B}_{av}$ which are the primitive fields [17, Sect. 29]). Hence, these materials can be characterized by the following constitutive relations (valid for generic bianisotropic linear media [24], [25]):

$$\mathbf{D} = \varepsilon_0 \overline{\overline{\varepsilon_r}} . \mathbf{E}_{av} + \sqrt{\varepsilon_0 \mu_0} \overline{\overline{\xi}} . \mathbf{H} \tag{3a}$$

$$\mathbf{B}_{av} = \sqrt{\varepsilon_0 \mu_0} \overline{\overline{\zeta}} . \mathbf{E}_{av} + \mu_0 \overline{\overline{\mu_r}} . \mathbf{H} \tag{3b}$$

where $\overline{\overline{\varepsilon_r}}(\omega)$ is the relative permittivity dyadic, $\overline{\overline{\mu_r}}(\omega)$ is the relative permeability, and $\overline{\overline{\xi}}(\omega)$ and $\overline{\overline{\zeta}}(\omega)$ are (dimensionless) parameters that characterize the magneto-electric coupling (gyrotropy [18]). Often, the terms $\overline{\overline{\xi}}(\omega)$ and $\overline{\overline{\zeta}}(\omega)$ are negligible, and so the material can be described using uniquely permittivity and permeability tensors. In fact, it is well-known that gyrotropy can only be observed in media that possess no center of symmetry. This is the case of some naturally occurring materials studied in crystal optics [18] and some artificial materials [23], [24].

It is important to underline that the above classic model is meaningful only when the approximation $\langle \mathbf{J}_d \rangle \approx j\omega \mathbf{P} + \nabla \times \mathbf{M}$ holds and the higher-order multipole moments can be neglected. Also it is implicitly assumed that the medium is local, i.e. that **D** and **H** at a given "point" of space can be written in terms of $\mathbf{E}_{av}$ and $\mathbf{B}_{av}$ at the same "point" of space, as implied by (3). Otherwise the medium is spatially dispersive [17]-[18]. For ordinary dielectrics, where the lattice constant ($a \sim 0.1 nm$) is several orders of magnitude smaller than the wavelength of radiation, these conditions are typically verified, and so the model (3) can be used to describe the propagation of waves. However, in common microstructured artificial materials the lattice constant is typically only 5-10 times larger than the wavelength of radiation, and so these effects may not be negligible [19]-[21], and the approximation (1) may not be accurate. As discussed in [17, Sect. 79], the same situation may also occur for natural media at the optical regime. Moreover, as argued in [17, Sect. 79], in natural media "the magnetic permeability ceases to have physical meaning at relatively low frequencies" and to take into account deviations of $\mu(\omega)$ from unity "would then be an unwarrantable refinement".

In general, in presence of spatial dispersion, the introduction of the effective permeability tensor $\overline{\overline{\mu}}_r$ (as well as of $\overline{\overline{\xi}}$ and $\overline{\overline{\zeta}}$) may not be meaningful [17, Sect. 79]. The problem is that the separation of the mean microscopic current as in (1) is not significant, because **P** and **M** may not be related with average fields through local relations. Due to this reason, as discussed in [17, Sect. 103], it is more appropriate to formulate the macroscopic Maxwell-Equations using alternative phenomenological constitutive relations. More specifically, it is preferable to include all the terms resulting from the averaging of the microscopic currents directly into the definition of the electric displacement **D**, without introducing a magnetization vector. Within this approach, as detailed below, the homogenized medium is described solely by a dielectric function of the form $\overline{\overline{\varepsilon}}_{eff}(\omega, \mathbf{k})$, where **k** is the wave vector. The effects of spatial dispersion have been studied in crystal optics, plasma physics and metal optics [18], and more recently in metamaterials [19], [21].

Following [17], [18], for a non-local medium we have the following definitions (compare with (2)):

$$\mathbf{D}_g = \varepsilon_0 \mathbf{E}_{av} + \mathbf{P}_g \qquad (4a)$$

$$\mathbf{H}_g = \frac{\mathbf{B}_{av}}{\mu_0} \qquad (4b)$$

where by definition $\mathbf{P}_g = \langle \mathbf{J}_d \rangle / j\omega$. We introduced the subscript "g" to underline that the electric displacement and the magnetic field defined as in (4) differ from the classical definition (2). In fact, as referred above, in this phenomenological model all the microscopic currents are included directly in the definition of the dielectric displacement. From (1), it is evident that $\mathbf{P}_g = \mathbf{P} + \nabla \times \mathbf{M} / j\omega + ....$ Thus, $\mathbf{P}_g$ is a generalized polarization vector that contains the effect of the dipolar moments, and in addition, the effect of all higher-order multipole moments. The constitutive relations corresponding to (4) are completely different from the constitutive relations (3). In fact, $\mathbf{D}_g$ cannot be related with the average field $\mathbf{E}_{av}$ through a local relation, since the polarization $\mathbf{P}_g$ at one point of space depends on the distribution of the macroscopic electric field in a neighborhood of the considered point. Due to this reason, it is commonly assumed that $\mathbf{D}_g$ can be written in terms of a convolution of $\mathbf{E}_{av}$ with another function [17, Sect. 103]. Thus, in the Fourier domain (dual of the **r**-domain), the following constitutive relations can be introduced:

$$\tilde{\mathbf{D}}_g = \overline{\overline{\varepsilon_{eff}}}(\omega, \mathbf{k}) . \tilde{\mathbf{E}}_{av} \qquad (5a)$$

$$\tilde{\mathbf{H}}_g = \frac{\tilde{\mathbf{B}}_{av}}{\mu_0} \qquad (5b)$$

where $\overline{\overline{\varepsilon_{eff}}}(\omega, \mathbf{k})$ is the dielectric function. In the above the symbol "~" represents the Fourier transform with respect to the spatial coordinates (the Fourier transform variable is the wave vector $\mathbf{k} = (k_x, k_y, k_z)$). Again, we stress that when the material is spatially dispersive the constitutive relations assume the simple form (5) only in the Fourier domain (or equivalently for plane wave solutions of Maxwell-Equations), because $\mathbf{P}_g$ cannot be locally related with $\mathbf{E}_{av}$. As referred before, in the spatial domain, the fundamental fields are related with the permittivity through a convolution [17]. For simplicity, in the rest of this paper we will drop the superscript "~", being implicit from the context if a given formula holds in the spatial domain, in the **k**-domain, or in both. Notice that when using (4) it is not necessary to introduce a magnetic permeability tensor: all the effects are directly taken into account by $\overline{\overline{\varepsilon}}_{eff}(\omega, \mathbf{k})$, including the effect of high order multipoles.

In this work we adopt the non-local phenomenological model based on (5). In section III, we will develop a microscopic theory to characterize arbitrary periodic non-magnetic metamaterials using the constitutive relations (5), and explain how the permittivity dyadic $\overline{\overline{\varepsilon}}_{eff}(\omega, \mathbf{k})$ can be numerically calculated. There are several advantages in extracting $\overline{\overline{\varepsilon}}_{eff}(\omega, \mathbf{k})$, rather than the parameters implicit in the bianisotropic model (3). The first advantage is that the non-local theory (5) is more general than the local theory (3), and consequently can be applied in more diverse situations/problems. Indeed, (3) should be regarded as a particular case of (5). The second advantage is that it is easier to formulate a microscopic theory to calculate $\overline{\overline{\varepsilon}}_{eff}(\omega, \mathbf{k})$, rather than a microscopic theory to directly calculate the parameters of the bianisotropic model. Also, very importantly, when the spatial dispersion is weak, it is always possible to extract the bianisotropic model parameters directly from $\overline{\overline{\varepsilon}}_{eff}(\omega, \mathbf{k})$ by expanding it in a Taylor series [17, Sect. 103-104], [25]. This topic is briefly discussed in what follows. An excellent discussion is also given in [26].

In fact, the key result (see [17], [26]) is that any (non-spatially dispersive) bianisotropic linear medium can always be described by the constitutive relations (5) (but the converse is not true). It is easy to verify that the relation between the parameters implicit in (3) and the equivalent $\overline{\overline{\varepsilon}}_{eff}$ is,

$$\frac{\overline{\overline{\varepsilon}}_{eff}}{\varepsilon_0}(\omega,\mathbf{k}) = \overline{\overline{\varepsilon}}_r - \overline{\overline{\xi}}.\overline{\overline{\mu}}_r^{-1}.\overline{\overline{\zeta}} + \left(\overline{\overline{\xi}}.\overline{\overline{\mu}}_r^{-1} \times \frac{\mathbf{k}}{\beta} - \frac{\mathbf{k}}{\beta} \times \overline{\overline{\mu}}_r^{-1}.\overline{\overline{\zeta}}\right) + \frac{\mathbf{k}}{\beta} \times \left(\overline{\overline{\mu}}_r^{-1} - \overline{\overline{\mathbf{I}}}\right) \times \frac{\mathbf{k}}{\beta} \quad (6)$$

where $\beta = \omega\sqrt{\varepsilon_0\mu_0}$ and $\overline{\overline{\mathbf{I}}}$ is the identity dyadic. Quite interestingly, the above formula also confirms that a material can be described using the bianisotropic model (3), only if the spatial dispersion is weak [17, Sect. 103-104], i.e. only if $\overline{\overline{\varepsilon}}_{eff}$ is a quadratic function of $\mathbf{k}$. Moreover, (6) demonstrates that the tensors $\overline{\overline{\xi}}$ and $\overline{\overline{\zeta}}$ (related with gyrotropy) can be determined from the first order derivatives of $\overline{\overline{\varepsilon}}_{eff}$ in $\mathbf{k}$ [17]-[18], and that the magnetic permeability is related with the second order derivatives of $\overline{\overline{\varepsilon}}_{eff}$ in $\mathbf{k}$. More details will be given in section V, when we discuss the numerical calculation of the magnetic permeability of metamaterials formed by split ring resonators. A more detailed analysis of this topic will be published elsewhere.

To conclude this section, we discuss which set of constitutive relations, (3) or (5), is better. More specifically, when a medium can be characterized by both (3) and (5) (with the corresponding effective parameters linked by (6)), which of the phenomenological constitutive relations is preferable? It is simple to verify that both sets of constitutive relations predict exactly the same dispersion characteristics $\omega = \omega(\mathbf{k})$ for plane waves, and also the same average fields $\mathbf{E}_{av}$ and $\mathbf{B}_{av}$. Despite that, in our opinion, the classical local constitutive relations (3) are more complete and powerful. The reason is very simple: while (3) is valid in the spatial domain, (5) is valid only in the Fourier domain. This remark shows that only the local constitutive relations (3) can be used to solve boundary value problems using the classical boundary conditions at an interface [26] (continuity of the tangential components of

$\mathbf{E}_{av}$ and $\mathbf{H}$). The non-local model (5) is of very limited utility for problems involving interfaces, because (5) only applies in the Fourier-space, and so implicitly assumes an infinite unbounded homogeneous structure. In fact, it is well-known that electromagnetic problems in spatially dispersive media are difficult to solve [18]. Even a simple plane wave incidence problem is not trivial because additional boundary conditions may be required at an interface [18], [27]. Thus, from that point of view, the classic model is more complete and powerful. Nevertheless, we underline again that the classic model (3) only applies when the spatial dispersion is weak, which may not be the case in some natural materials and certainly in many metamaterials.

## III. THE HOMOGENIZATION TECHNIQUE

Here, we develop a microscopic theory that enables the calculation of the non-local dielectric function $\overline{\overline{\varepsilon}}_{eff}(\omega, \mathbf{k})$ for generic non-magnetic periodic materials. The physical motivation and mathematical foundation of the proposed extraction method is presented in section III.A. Then, in section III.B it is shown that even though the mathematical formulation of section III.A is simple and intuitive, the extraction problem is not well defined when $(\omega, \mathbf{k})$ are associated with an electromagnetic mode of the periodic medium. To circumvent this drawback, we regularize the mathematical problem proving that it can be reformulated as integral-differential system. To conclude this section, we describe some elementary properties of the dielectric function $\overline{\overline{\varepsilon}}_{eff}(\omega, \mathbf{k})$.

### A. *Microscopic theory*

In order to retrieve the effective parameters of a periodic medium, our idea is to excite the structure with a periodic source that imposes a desired phase modulation in the unit cell. The unknown dielectric function can then be computed from the induced microscopic currents.

Notice that a similar procedure is used for example in plasma physics to characterize the wave vector dependence of the dielectric function of an electron gas [28, pp. 280].

To better explain the proposed concepts, consider the frequency dependent Maxwell-Equations in a non-magnetic medium,

$$\nabla \times \mathbf{E} = -j\omega \mathbf{B} \tag{7a}$$

$$\nabla \times \frac{\mathbf{B}}{\mu_0} = \mathbf{J}_e + \varepsilon_0 \varepsilon_r j\omega \mathbf{E} \tag{7b}$$

where $\mathbf{E}$ is the microscopic electric field, and $\mathbf{B}$ is the microscopic induction field, and $\mathbf{J}_e$ is the applied electric current density (source of fields). To calculate $\overline{\overline{\varepsilon_{eff}}}$ for given $(\omega, \mathbf{k})$, we impose that $\mathbf{J}_e$ has the Floquet property, i.e. $\mathbf{J}_e \exp(j\mathbf{k}.\mathbf{r})$ is periodic in the crystal. An immediate consequence is that the solution $(\mathbf{E}, \mathbf{B})$ of (7) also has the Floquet property. For convenience we rewrite (7b) as follows,

$$\nabla \times \frac{\mathbf{B}}{\mu_0} = \mathbf{J}_e + \mathbf{J}_d + \varepsilon_0 j\omega \mathbf{E} \tag{7b'}$$

where $\mathbf{J}_d = \varepsilon_0 (\varepsilon_r - 1) j\omega \mathbf{E}$ is the induced current relative to the host medium (assumed vacuum for simplicity), which is regarded here as the microscopic current induced in the host material, consistently with the definition of the previous section.

In order to obtain the counterpart of (7) in the Fourier space, we define the average (macroscopic) fields $\mathbf{E}_{av}$ and $\mathbf{B}_{av}$ as follows,

$$\mathbf{E}_{av} = \frac{1}{V_{cell}} \int_\Omega \mathbf{E}(\mathbf{r}) e^{+j\mathbf{k}.\mathbf{r}} d^3\mathbf{r}, \qquad \mathbf{B}_{av} = \frac{1}{V_{cell}} \int_\Omega \mathbf{B}(\mathbf{r}) e^{+j\mathbf{k}.\mathbf{r}} d^3\mathbf{r} \tag{8}$$

It can be easily verified that the following equations hold exactly,

$$-\mathbf{k} \times \mathbf{E}_{av} + \omega \mathbf{B}_{av} = 0 \tag{9a}$$

$$\omega(\varepsilon_0 \mathbf{E}_{av} + \mathbf{P}_g) + \mathbf{k} \times \frac{\mathbf{B}_{av}}{\mu_0} = -\omega \mathbf{P}_e \tag{9b}$$

where $\mathbf{P}_e = \frac{1}{V_{cell} j\omega} \int_\Omega \mathbf{J}_e e^{+j\mathbf{k}.\mathbf{r}} d^3\mathbf{r}$ is the *applied* polarization vector, and $\mathbf{P}_g$ is the *generalized* (induced) polarization vector,

$$\mathbf{P}_g = \frac{1}{V_{cell} j\omega} \int_\Omega \mathbf{J}_d e^{+j\mathbf{k}\cdot\mathbf{r}} d^3\mathbf{r} \tag{10}$$

As referred in section II, $\mathbf{P}_g$ is closely related to the classic polarization vector. Indeed, if the exponential inside the integral is expanded in powers of the argument, the leading term corresponds exactly to the standard polarization vector (i.e. the average electric dipole moment in a unit cell; the polarization is relative to the host medium). The higher order terms can be related to the magnetization vector and other multipole moments.

For future reference, we notice that when the unit cell contains perfectly electric conducting (PEC) surfaces, the polarization vector becomes:

$$\mathbf{P}_g = \frac{1}{V_{cell} j\omega} \left( \int_{\partial D} \mathbf{J}_c e^{+j\mathbf{k}\cdot\mathbf{r}} ds + \int_\Omega \mathbf{J}_d e^{+j\mathbf{k}\cdot\mathbf{r}} d^3\mathbf{r} \right) \tag{11}$$

where $\partial D$ is the PEC surface, and $\mathbf{J}_c = \hat{\mathbf{v}} \times [\mathbf{B}/\mu_0]$ is the corresponding surface current (see Fig. 1).

Consistently with (4) and (5), the dielectric function $\overline{\overline{\varepsilon}}_{eff} = \overline{\overline{\varepsilon}}_{eff}(\omega, \mathbf{k})$ is defined in such a way that:

$$\overline{\overline{\varepsilon}}_{eff} \cdot \mathbf{E}_{av} = \varepsilon_0 \mathbf{E}_{av} + \mathbf{P}_g \tag{12}$$

Notice that for fixed $(\omega, \mathbf{k})$ the dielectric function $\overline{\overline{\varepsilon}}_{eff}$ can be completely determined from (12) provided $\mathbf{P}_g$ is known for three independent vectors $\mathbf{E}_{av}$ (e.g. for $\mathbf{E}_{av} \sim \hat{\mathbf{u}}_i$ where $\hat{\mathbf{u}}_i$ is directed along the coordinate axes). Thus, the recipe that we propose here to calculate the dielectric function can be summarized as follows. First, for fixed $(\omega, \mathbf{k})$, select three different distributions for the applied current $\mathbf{J}_e$ such that the corresponding induced average fields $\mathbf{E}_{av}$ form an independent set of vectors with dimension 3. Then, solve the source driven electromagnetic problem (7) to obtain the microscopic fields. From the microscopic fields calculate $\mathbf{E}_{av}$ and $\mathbf{P}_g$, and then, using (12), obtain the desired dielectric function $\overline{\overline{\varepsilon}}_{eff}$. In section III.B, it is explained in details how the outlined method can be applied in practice.

We should note however that the proposed homogenization procedure is not completely self-consistent. Indeed, the calculated $\overline{\overline{\varepsilon}}_{eff}$ will depend to some extent on the specific spatial variation of the chosen applied currents $\mathbf{J}_e$, i.e. depending on the specific choice for $\mathbf{J}_e$ the computed dielectric function may not be the same. Nevertheless, for low frequencies, when the dimensions of the cell are much smaller than the wavelength, it is reasonable to expect that the dependence of the induced polarization $\mathbf{P}_g$ on the specific spatial distribution of $\mathbf{J}_e$ is relatively weak. Hence, we propose to calculate the effective parameters by computing $\mathbf{P}_g$ under the excitation of a spatially uniform density of current, i.e. we assume that $\mathbf{J}_e$ is of the form:

$$\mathbf{J}_e = \mathbf{J}_{e,av} e^{-j\mathbf{k}\cdot\mathbf{r}} \tag{13}$$

where $\mathbf{J}_{e,av}$ is a constant vector, i.e. independent of $\mathbf{r}$. As referred before, $\mathbf{J}_e$ has the Floquet property. We prove in section III.B that (13) allows formulating a self-consistent homogenization theory. More physical and mathematical insights, and an alternative regularized formulation and integral equation based solution will also be described.

## B. Regularized Integral-differential formulation

The straightforward homogenization approach described before is attractive due to its simplicity. Notice that the approach is not based, by any means, on the solution of an eigenvalue problem. In fact, the proposed homogenization approach is a source-driven problem: for a given $\mathbf{J}_e$ we can solve the Maxwell-Equations (7) in the periodic medium, and afterwards compute the effective parameters of the structure using (12). This is very convenient, because it allows calculating the effective parameters even when materials are lossy or in frequency bands lying in a complete band-gap.

There is however an inconvenience with the straightforward formulation presented in section III.A. Indeed, when $(\omega, \mathbf{k})$ is associated with an electromagnetic mode of the periodic medium, in general (7) does not have a solution because the corresponding homogeneous system (with $\mathbf{J}_e = 0$) has a non-trivial solution. In fact, it is a well-known result that a non-homogeneous linear equation is solvable only if the independent term (the source) is orthogonal to the solutions of the homogeneous adjoint equation. Hence, when $(\omega, \mathbf{k})$ is associated with a modal solution, equation (7) can be solved only for very specific $\mathbf{J}_e$. The physical reason for this result is that if the medium is excited with a source with the same $(\omega, \mathbf{k})$ as an eigenmode, then a resonance is hit and the amplitude of the fields may grow without limit.

This is an undesired property because the effective parameters of a composite medium are intrinsically related with its electromagnetic modes. To circumvent this drawback, next we derive an alternative regularized formulation for the homogenization problem. The idea behind this regularized approach is that we can tune the amplitude of the imposed current $\mathbf{J}_e$ in such a way to enforce that the induced microscopic electric field has a given average value $\mathbf{E}_{av}$, guaranteeing in this way that a resonance is not hit when $(\omega, \mathbf{k})$ is associated with an eigenmode. This is possible because the amplitude of $\mathbf{J}_e$ becomes interrelated with the induced microscopic currents in the periodic medium, in such a way that depolarization effects prevent the fields in the medium to grow without limit when a resonance is approached. The details of the method are presented next.

To begin with, we will relate the amplitude of $\mathbf{J}_e$ with the induced macroscopic field $\mathbf{E}_{av}$. To this end, we use (9) to obtain after straightforward manipulations that,

$$\frac{\mathbf{P}_e}{\varepsilon_0} = \frac{1}{\beta^2} \frac{1}{V_{cell}} \overline{\overline{\mathbf{G}}}_0^{-1} . \mathbf{E}_{av} - \frac{\mathbf{P}_g}{\varepsilon_0} \tag{14}$$

where we defined the dyadic $\overline{\overline{\mathbf{G}}}_0$ and the respective inverse $\overline{\overline{\mathbf{G}}}_0^{-1}$ by,

$$\overline{\overline{\mathbf{G}}}_0 = \frac{1}{V_{cell}} \frac{1}{\beta^2} \frac{\beta^2 \overline{\overline{\mathbf{I}}} - \mathbf{kk}}{k^2 - \beta^2} \tag{15a}$$

$$\overline{\overline{\mathbf{G}}}_0^{-1} = -V_{cell}\left((\beta^2 - k^2)\overline{\overline{\mathbf{I}}} + \mathbf{kk}\right) \tag{15b}$$

and $k^2 = \mathbf{k}.\mathbf{k}$ and $\beta = \omega\sqrt{\varepsilon_0\mu_0}$ is the wave number in the host medium.

Equation (14) relates the applied polarization vector, with the average electric field and the induced polarization vector. It is convenient to rewrite the equation in terms of two auxiliary operators $\hat{\mathbf{P}}$ and $\hat{\mathbf{P}}_{av}$, defined next. The polarization operator, $\hat{\mathbf{P}}$, transforms the electric vector field into the corresponding (generalized) polarization vector, $\hat{\mathbf{P}}: \mathbf{E} \to \mathbf{P} = \hat{\mathbf{P}}(\mathbf{E})$, where:

$$\frac{\hat{\mathbf{P}}(\mathbf{E})}{\varepsilon_0} = \frac{1}{V_{cell}}\left(\frac{1}{\beta^2}\int_{\partial D}\hat{\mathbf{v}}\times[\nabla\times\mathbf{E}]e^{+j\mathbf{k}.\mathbf{r}}ds + \int_{\Omega}(\varepsilon_r - 1)\mathbf{E}\,e^{+j\mathbf{k}.\mathbf{r}}d^3\mathbf{r}\right) \tag{16}$$

In the above $[\nabla\times\mathbf{E}] = \nabla\times\mathbf{E}^+ - \nabla\times\mathbf{E}^-$ stands for the discontinuity of the curl of $\mathbf{E}$ at the metallic surfaces, and $\nabla\times\mathbf{E}^+$ is evaluated at the outer side of $\partial D$ (see Fig. 1). It can be easily verified that the above definition is consistent with (11). The second operator, $\hat{\mathbf{P}}_{av}$, acts on constant vectors (not on vector fields), $\hat{\mathbf{P}}_{av}: \mathbf{E}_{av} \to \hat{\mathbf{P}}_{av}(\mathbf{E}_{av})$, and is given by:

$$\frac{\hat{\mathbf{P}}_{av}(\mathbf{E}_{av})}{\varepsilon_0} = \frac{1}{\beta^2}\frac{1}{V_{cell}}\overline{\overline{\mathbf{G}}}_0^{-1}.\mathbf{E}_{av} \tag{17}$$

Equation (14) is thus equivalent to:

$$\mathbf{P}_e = \hat{\mathbf{P}}_{av}(\mathbf{E}_{av}) - \hat{\mathbf{P}}(\mathbf{E}) \tag{18}$$

Using the definition of $\mathbf{P}_e$ and (13), we find that applied density of current is such that:

$$\mathbf{J}_e = j\omega\left(\hat{\mathbf{P}}_{av}(\mathbf{E}_{av}) - \hat{\mathbf{P}}(\mathbf{E})\right)e^{-j\mathbf{k}.\mathbf{r}} \tag{19}$$

We can use this formula to regularize the proposed homogenization approach. Thus, instead of imposing a fixed amplitude for $\mathbf{J}_e$ as in the straightforward formulation of section III.A, we can write $\mathbf{J}_e$ as a function of the macroscopic field $\mathbf{E}_{av}$, and consequently also as a function of the microscopic field $\mathbf{E}$, i.e. $\mathbf{J}_e$, which is source of the microscopic fields, becomes also itself a function of the microscopic fields. This feedback mechanism prevents a resonance from being excited when $(\omega, \mathbf{k})$ is associated with an electromagnetic mode.

To clarify this aspect, we substitute (19) in (7) to obtain:

$$\nabla \times \mathbf{E} = -j\omega \mathbf{B} \qquad (20a)$$

$$\nabla \times \frac{\mathbf{B}}{\mu_0} = j\omega \left( \hat{\mathbf{P}}_{av} (\mathbf{E}_{av}) - \hat{\mathbf{P}}(\mathbf{E}) \right) e^{-j\mathbf{k}\cdot\mathbf{r}} + \varepsilon_0 \varepsilon_r j\omega \mathbf{E} \qquad (20b)$$

Even though the above system is closely related with (7), there are some important and relevant differences. First of all, unlike (7), system (20) is an integral differential-system, i.e. both differential operators ($\nabla \times$) and integral operators ($\hat{\mathbf{P}}(.)$) act on the electromagnetic fields. Note that $\hat{\mathbf{P}}(.)$ yields the generalized polarization of the unknown field $\mathbf{E}$, which involves the integration of the electric field over the unit cell.

Another key property is that while in (7) the source of fields is $\mathbf{J}_e$, in (20) the source of fields (from a mathematical point of view) is the constant vector $\mathbf{E}_{av}$. An important consequence is that the solutions of the homogeneous problem ($\mathbf{J}_e = 0$) associated with (7) are different from the solutions of the homogeneous system ($\mathbf{E}_{av} = 0$) associated with (20), i.e. the systems have different null spaces. In particular, the electromagnetic modes of the periodic medium are associated with a non-trivial $\mathbf{E}_{av}$ and so do not belong to the null space of (20). Thus, unlike in (7) the formulation based on (20), can be used to compute the effective parameters of the composite medium even if $(\omega, \mathbf{k})$ is associated with an electromagnetic mode. In fact, when $(\omega, \mathbf{k})$ is associated with a modal solution, we have that $\hat{\mathbf{P}}_{av}(\mathbf{E}_{av}) = \hat{\mathbf{P}}(\mathbf{E})$ and thus the amplitude of the imposed current in (20) vanishes, avoiding the resonance from being hit. However, since $\mathbf{E}_{av}$ is different from zero (20) is still a well-formulated source driven problem.

Note that the effective parameters retrieved by solving (7) are exactly the same as those obtained by solving (20). We underline that the only difference between the two formulations is that the regularized formulation can be applied even when $(\omega, \mathbf{k})$ is associated with an electromagnetic mode.

Another interesting property of the integral-differential system (20), is that its solution $(\mathbf{E}, \mathbf{B})$ for a given $\mathbf{E}_{av}$ is such that the spatial average of $\mathbf{E}$ is precisely $\mathbf{E}_{av}$. The proof of this result is a direct consequence of (19).

The possible downside of the regularized approach is that an integral-differential system is in general more difficult to solve than a differential system. Nevertheless, in section IV we will prove that (20) can be reduced to a standard integral equation, which can be solved numerically using well-known techniques.

## C. The characteristic equation

In this section, we confirm that the dielectric function defined by (12) can predict the relevant properties of the macroscopic electromagnetic modes of the periodic medium. In addition, we briefly review the properties of the associated dispersion equation and relate the polarization of the macroscopic field with the dielectric function.

To begin with, we note that the applied polarization vector $\mathbf{P}_e$ can be zero (for a non-trivial $\mathbf{E}_{av}$, see (18)) if and only if $(\omega, \mathbf{k})$ is associated with an electromagnetic mode. Hence, substituting (12) in (9), we conclude that the homogeneous system,

$$-\mathbf{k} \times \mathbf{E}_{av} + \omega \mathbf{B}_{av} = 0 \tag{21a}$$

$$\omega \overline{\overline{\varepsilon_{eff}}} \cdot \mathbf{E}_{av} + \mathbf{k} \times \frac{\mathbf{B}_{av}}{\mu_0} = 0 \tag{21b}$$

has a non-trivial solution if and only if $(\omega, \mathbf{k})$ is associated with an electromagnetic mode. This is an exact result valid for arbitrary $(\omega, \mathbf{k})$, not necessarily in the long wavelength limit. The system (21) is precisely the same as that obtained for plane wave solutions in a homogeneous non-magnetic anisotropic medium characterized by the permittivity $\overline{\overline{\varepsilon}}_{eff}$. This proves that the dielectric function defined by (12) can indeed be used to obtain the band structure/average fields of an arbitrary electromagnetic mode.

For the sake of completeness, next we briefly review the properties of the characteristic system (21). As is well known [18, pp. 25], (21) implies that the average electric field satisfies the characteristic system:

$$\left( \frac{\overline{\overline{\varepsilon}}_{eff}}{\varepsilon_0} + \frac{1}{\beta^2} \mathbf{k}\mathbf{k} - \frac{k^2}{\beta^2} \overline{\overline{\mathbf{I}}} \right) . \mathbf{E}_{av} = \mathbf{0} \tag{22}$$

From (21b) we also have that $\mathbf{k}.\overline{\overline{\varepsilon}}_{eff}.\mathbf{E}_{av} = 0$. On the other hand, manipulating (22) we easily find that:

$$\mathbf{E}_{av} = -(\mathbf{k}.\mathbf{E}_{av}) \left( \beta^2 \frac{\overline{\overline{\varepsilon}}_{eff}}{\varepsilon_0} - k^2 \overline{\overline{\mathbf{I}}} \right)^{-1} .\mathbf{k} \tag{23}$$

Thus, provided the average field is not transverse, i.e. provided $\mathbf{k}.\mathbf{E}_{av} \neq 0$, we conclude that the macroscopic electric field can be written as:

$$\mathbf{E}_{av} \propto \left( \frac{\overline{\overline{\varepsilon}}_{eff}}{\varepsilon_0} - \frac{k^2}{\beta^2} \overline{\overline{\mathbf{I}}} \right)^{-1} .\frac{\mathbf{k}}{\beta}, \qquad \text{if } \mathbf{k}.\mathbf{E}_{av} \neq 0 \tag{24}$$

Also, multiplying both sides of (23) by $\mathbf{k}$, we readily find that $(\omega, \mathbf{k})$ satisfies the characteristic equation:

$$-1 = \mathbf{k}. \left( \beta^2 \frac{\overline{\overline{\varepsilon}}_{eff}}{\varepsilon_0} - k^2 \overline{\overline{\mathbf{I}}} \right)^{-1} .\mathbf{k} \qquad \text{if } \mathbf{k}.\mathbf{E}_{av} \neq 0 \tag{25}$$

Note that whenever $\mathbf{k}.\mathbf{E}_{av} = 0$ the dyadic in the above equation must have a singularity. The solutions, $\omega = \omega(\mathbf{k})$, of (25) yield the dispersion of the electromagnetic modes.

### D. Properties of the permittivity dyadic

Some important properties of the dielectric function defined as in (12) are enunciated next. The proof of the results can be found in Appendix A. Note that the derived properties are consistent with [17, Sect. 103]. Below the superscript "$t$" refers to the transpose dyadic.

- **P1)** $\overline{\overline{\varepsilon}}_{eff}(\omega, \mathbf{k}) = \overline{\overline{\varepsilon}}_{eff}^{\,t}(\omega, -\mathbf{k})$.

- **P2)** Let $\mathbf{T}$ be a translation. Let us suppose that a given metamaterial is characterized by the dielectric function $\overline{\overline{\varepsilon}}_{eff}$, and that the metamaterial resulting from the application of $\mathbf{T}$ to the original structure is characterized by the dielectric function $\overline{\overline{\varepsilon}}'_{eff}$. Then, $\overline{\overline{\varepsilon}}'_{eff}(\omega, \mathbf{k}) = \overline{\overline{\varepsilon}}_{eff}(\omega, \mathbf{k})$. In particular, the definition of the dielectric function is independent of the origin of the coordinate system.

- **P3)** Let $\mathbf{S}$ be an isometry (a rotation or a reflection): $\mathbf{S}.\mathbf{S}^t = \overline{\overline{\mathbf{I}}}$. Let us suppose that a given metamaterial is characterized by the dielectric function $\overline{\overline{\varepsilon}}_{eff}$, and that the metamaterial resulting from the application of $\mathbf{S}$ to the original structure is characterized by the dielectric function $\overline{\overline{\varepsilon}}'_{eff}$. Then, $\overline{\overline{\varepsilon}}'_{eff}(\omega, \mathbf{S}.\mathbf{k}) = \mathbf{S}.\overline{\overline{\varepsilon}}_{eff}(\omega, \mathbf{k}).\mathbf{S}^t$.

- **P4)** If a metamaterial is invariant to an affine isometry $\mathbf{T} \circ \mathbf{S}$ (rotation/reflection followed by a translation) then its dielectric function satisfies $\overline{\overline{\varepsilon}}_{eff}(\omega, \mathbf{S}.\mathbf{k}) = \mathbf{S}.\overline{\overline{\varepsilon}}_{eff}(\omega, \mathbf{k}).\mathbf{S}^t$. In particular, if the material has a center of symmetry, i.e. there is a point such that material is invariant to the transformation $\mathbf{S}: \mathbf{r} \to -\mathbf{r}$, then $\overline{\overline{\varepsilon}}_{eff}(\omega, \mathbf{k}) = \overline{\overline{\varepsilon}}_{eff}(\omega, -\mathbf{k})$.

## IV. INTEGRAL EQUATION FORMULATION

So far the emphasis of this work has been on the theoretical aspects and properties of the homogenization problem. As discussed in section III, the regularized homogenization problem is an integral-differential system. In what follows, we will prove that this complex system can be reduced to a standard integral equation, which can be solved using known techniques. This paves the way so that the proposed approach can be used to solve practical problems using numerical methods, as exemplified in section V.

## A. *Integral representation of the electric field*

Here, we prove that microscopic electric field – solution of the homogenization problem (20) – has the integral representation (it is assumed that $\mathbf{r} \notin \partial D$):

$$\mathbf{E}(\mathbf{r}) = \mathbf{E}_{av} e^{-j\mathbf{k}.\mathbf{r}} + \int_{\partial D} \overline{\overline{\mathbf{G}_{p0}}}(\mathbf{r}|\mathbf{r}') \cdot (\hat{\mathbf{v}}' \times [\nabla' \times \mathbf{E}]) ds' + \int_{\Omega} \overline{\overline{\mathbf{G}_{p0}}}(\mathbf{r}|\mathbf{r}') \cdot \beta^2 (\varepsilon_r - 1) \mathbf{E} \, d^3\mathbf{r}' \quad (26)$$

where $\overline{\overline{\mathbf{G}_{p0}}}$ is a Green function dyadic introduced below. The integral representation establishes that the microscopic field $\mathbf{E}$ can be written in terms of the induced microscopic currents and of the macroscopic electric field. This important result is used in section IV.B to reduce the homogenization problem to an integral equation.

The proof of (26) is delineated next. To begin with, we introduce the lattice Green function $\Phi_p = \Phi_p(\mathbf{r}|\mathbf{r}'; \beta, \mathbf{k})$ [29], [30], which verifies:

$$\nabla^2 \Phi_p + \beta^2 \Phi_p = -\sum_{\mathbf{I}} \delta(\mathbf{r} - \mathbf{r}' - \mathbf{r}_{\mathbf{I}}) e^{-j\mathbf{k}.(\mathbf{r}-\mathbf{r}')} \quad (27)$$

where $\mathbf{I} = (i_1, i_2, i_3)$ is a multi-index of integers, $\mathbf{r} = (x, y, z)$ is the observation point, $\mathbf{r}' = (x', y', z')$ is a source point, $\mathbf{r}_{\mathbf{I}} = i_1 \mathbf{a}_1 + i_2 \mathbf{a}_2 + i_3 \mathbf{a}_3$ is a lattice point, and $\delta$ is Dirac's distribution. The Green function depends on both $\beta = \omega \sqrt{\varepsilon_0 \mu_0}$ and $\mathbf{k}$.

The lattice Green function has the spectral representation:

$$\Phi_p(\mathbf{r}|\mathbf{r}') = \frac{1}{V_{cell}} \sum_{\mathbf{J}} \frac{e^{-j\mathbf{k}_{\mathbf{J}}.(\mathbf{r}-\mathbf{r}')}}{\mathbf{k}_{\mathbf{J}}.\mathbf{k}_{\mathbf{J}} - \beta^2}, \qquad \mathbf{k}_{\mathbf{J}} = \mathbf{k} + \mathbf{k}_{\mathbf{J}}^0 \quad (28)$$

where $V_{cell} = |\mathbf{a}_1 \cdot (\mathbf{a}_2 \times \mathbf{a}_3)|$ is the volume of the unit cell, $\mathbf{J} = (j_1, j_2, j_3)$ is a multi-index of integers, and $\mathbf{k}_{\mathbf{J}}^0 = j_1 \mathbf{b}_1 + j_2 \mathbf{b}_2 + j_3 \mathbf{b}_3$. The reciprocal lattice primitive vectors, $\mathbf{b}_n$, are implicitly defined by the relations $\mathbf{a}_m \cdot \mathbf{b}_n = 2\pi \delta_{m,n}$, $m,n=1,2,3$. Although the spectral representation is conceptually appealing due to its simplicity, the corresponding numerical series converges slowly. Fortunately, other representations with Gaussian and exponential convergence rates are also available [29, 30].

We also need to introduce the lattice Green dyadic, $\overline{\overline{\mathbf{G}}}_p = \overline{\overline{\mathbf{G}}}_p(\mathbf{r}|\mathbf{r}';\beta,\mathbf{k})$, which is the solution of:

$$\nabla \times \nabla \times \overline{\overline{\mathbf{G}}}_p - \beta^2 \overline{\overline{\mathbf{G}}}_p = \overline{\overline{\mathbf{I}}} e^{-j\mathbf{k}.(\mathbf{r}-\mathbf{r}')} \left( \sum_{\mathbf{I}} \delta(\mathbf{r}-\mathbf{r}'-\mathbf{r}_{\mathbf{I}}) \right) \tag{29}$$

It is straightforward to verify that:

$$\overline{\overline{\mathbf{G}}}_p = \left( \overline{\overline{\mathbf{I}}} + \frac{1}{\beta^2} \nabla \nabla \right) \Phi_p \tag{30a}$$

$$\overline{\overline{\mathbf{G}}}_p(\mathbf{r}|\mathbf{r}';\beta,\mathbf{k}) = \overline{\overline{\mathbf{G}}}_p^{\,t}(\mathbf{r}'|\mathbf{r};\beta,-\mathbf{k}) \tag{30b}$$

Using now (20), (29) and (30b), and standard integral equation techniques [31], it is possible to verify that if $\mathbf{E}$ is the solution of the homogenization problem associated with $(\mathbf{k},\omega,\mathbf{E}_{av})$, then,

$$\mathbf{E}(\mathbf{r}) = \int_{\partial D} \overline{\overline{\mathbf{G}}}_p(\mathbf{r}|\mathbf{r}').(\hat{\mathbf{v}}' \times [\nabla' \times \mathbf{E}])ds' + \int_{\Omega} \overline{\overline{\mathbf{G}}}_p(\mathbf{r}|\mathbf{r}').\beta^2(\varepsilon_r - 1)\mathbf{E}\, d^3\mathbf{r}' +$$
$$+ \left( \int_{\Omega} \overline{\overline{\mathbf{G}}}_p(\mathbf{r}|\mathbf{r}') e^{-j\mathbf{k}.\mathbf{r}'}\, d^3\mathbf{r}' \right).\beta^2 \left( \frac{\hat{\mathbf{P}}_{av}(\mathbf{E}_{av})}{\varepsilon_0} - \frac{\hat{\mathbf{P}}(\mathbf{E})}{\varepsilon_0} \right) \tag{31}$$

where the Green dyadic is associated with $(\beta,\mathbf{k})$.

Substituting (28) into (30a), we find that:

$$\int_{\Omega} \overline{\overline{\mathbf{G}}}_p(\mathbf{r}|\mathbf{r}') e^{-j\mathbf{k}.\mathbf{r}'}\, d^3\mathbf{r}' = V_{cell} \overline{\overline{\mathbf{G}}}_0 e^{-j\mathbf{k}.\mathbf{r}} \tag{32}$$

where $\overline{\overline{\mathbf{G}}}_0$ is given by (15a). It is also convenient to introduce the dyadic $\overline{\overline{\mathbf{G}}}_{p0} = \overline{\overline{\mathbf{G}}}_{p0}(\mathbf{r}|\mathbf{r}';\beta,\mathbf{k})$ defined by:

$$\overline{\overline{\mathbf{G}}}_{p0}(\mathbf{r}|\mathbf{r}') = \overline{\overline{\mathbf{G}}}_p(\mathbf{r}|\mathbf{r}') - \overline{\overline{\mathbf{G}}}_0 e^{-j\mathbf{k}.(\mathbf{r}-\mathbf{r}')} \tag{33}$$

Using (16), (17), and the previous formulas in (31), we conclude that $\mathbf{E}$ has the integral representation (26), as we wanted to prove.

Since the kernel of the integral in (26) is $\overline{\overline{\mathbf{G}}}_{p0}$, it is worth noting that it is the solution of,

$$\nabla \times \nabla \times \overline{\overline{\mathbf{G}}}_{p0} - \beta^2 \overline{\overline{\mathbf{G}}}_{p0} = \overline{\overline{\mathbf{I}}} e^{-j\mathbf{k}.(\mathbf{r}-\mathbf{r}')} \left( \sum_{\mathbf{I}} \delta(\mathbf{r} - \mathbf{r}' - \mathbf{r}_{\mathbf{I}}) - \frac{1}{V_{cell}} \right) \quad (34)$$

The dyadic $\overline{\overline{\mathbf{G}}}_{p0}$ can also be written as:

$$\overline{\overline{\mathbf{G}}}_{p0} = \left( \overline{\overline{\mathbf{I}}} + \frac{1}{\beta^2} \nabla \nabla \right) \Phi_{p0} \quad (35a)$$

$$\Phi_{p0}(\mathbf{r}|\mathbf{r}') = \Phi_{p}(\mathbf{r}|\mathbf{r}') - \frac{1}{V_{cell}} \frac{e^{-j\mathbf{k}.(\mathbf{r}-\mathbf{r}')}}{k^2 - \beta^2} = \frac{1}{V_{cell}} \sum_{\mathbf{J} \neq 0} \frac{e^{-j\mathbf{k_J}.(\mathbf{r}-\mathbf{r}')}}{\mathbf{k_J}.\mathbf{k_J} - \beta^2} \quad (35b)$$

Note that $\Phi_{p0}$ can be numerically evaluated very efficiently using the method proposed in [14, 29].

### B. Integral equation formulation

Here, (26) is used to demonstrate that the homogenization problem can be reduced to a standard integral equation. Using this result, we derive a closed form solution for the unknown dielectric function in terms of the inverse of an infinite matrix.

The integral representation (26) can be regarded as an integral equation with unknowns given by the microscopic currents $\mathbf{J}_d = \varepsilon_0 (\varepsilon_r - 1) j\omega \mathbf{E}$ at the dielectric inclusions, $\mathbf{J}_c = \hat{\mathbf{v}} \times [\mathbf{B}/\mu_0]$ at the PEC surfaces. For a given $\mathbf{E}_{av}$, the integral equation is obtained by imposing that (26) is verified in the inclusions, and that the tangential component of the electric field vanishes at the PEC surfaces. The integral equation can be discretized and numerically solved using standard techniques. In what follows, we discretize the integral equation using the method of moments (MoM) [31] assuming that the periodic material is purely dielectric. The case of PEC inclusions will be briefly described later.

We take the vector field $\mathbf{f} = (\varepsilon_r - 1)\mathbf{E}$, as the unknown of the integral equation. Notice that the vector density $\mathbf{f}$ vanishes in the host medium and is proportional to the polarization current $\mathbf{J}_d$. Supposing there are no PEC surfaces, and enforcing that (26) holds outside the host medium, we obtain the integral equation:

$$\frac{\mathbf{f}(\mathbf{r})}{(\varepsilon_r - 1)} = \mathbf{E}_{av} e^{-j\mathbf{k}\cdot\mathbf{r}} + \int_\Omega \overline{\overline{\mathbf{G}_{p0}}}(\mathbf{r}|\mathbf{r}') \cdot \beta^2 \mathbf{f}(\mathbf{r}') \, d^3\mathbf{r}' \qquad (36)$$

The identity is valid over $\{\mathbf{r} : \varepsilon_r(\mathbf{r}) - 1 \neq 0\}$, i.e. in the dielectric support of the inclusions. Note that because of the proposed regularization method, the kernel of the above integral equation $\overline{\overline{\mathbf{G}_{p0}}}(\mathbf{r}|\mathbf{r}')$, differs from the kernel $\overline{\overline{\mathbf{G}_{p}}}(\mathbf{r}|\mathbf{r}')$ associated with an integral representation of a generic electromagnetic mode. This is the reason why the MoM formulation proposed here is, in fact, numerically stable, even if $(\omega, \mathbf{k})$ is associated with modal solution. For a given $\mathbf{E}_{av}$ we can solve the integral equation (36) with respect to $\mathbf{f}$, as explained next.

First, we expand $\mathbf{f}$ in terms of the vector expansion functions $\mathbf{w}_1$, $\mathbf{w}_2$, ...:

$$\mathbf{f} = \sum_n c_n \mathbf{w}_n \qquad (37)$$

The set of expansion functions is assumed complete. Notice that $\mathbf{f}$ is obviously a Floquet field, i.e. $\mathbf{f} \exp(j\mathbf{k}\cdot\mathbf{r})$ is periodic. Thus, in general, the expansion functions are Floquet fields and, therefore, must depend explicitly on $\mathbf{k}$, i.e. $\mathbf{w}_n = \mathbf{w}_{n,\mathbf{k}}(\mathbf{r})$. The dependence on $\mathbf{k}$ can be suppressed only if the inclusions are non-connected (i.e. when the set $\{\mathbf{r} \in \Omega : \varepsilon_r(\mathbf{r}) - 1 \neq 0\}$ does not have two equivalent points spaced by a lattice primitive vector). For example, if the medium is formed by non-connected spheres the dependence on $\mathbf{k}$ may be suppressed, while if it is formed by continuous tubes (cylinders) the dependence must be implicitly incorporated in the expansion functions.

In order to reduce (36) to a linear system, we use the standard procedure of multiplying both sides of the equation with *test* functions and integrating the resulting expression over the unit cell. For simplicity we take the test functions equal to the expansion functions, with one difference: the test functions are of the form $\mathbf{w}_m = \mathbf{w}_{m,-\mathbf{k}}(\mathbf{r})$, i.e. are associated to the wave vector $-\mathbf{k}$ in order to "kill" the phase variation and improve numerical convergence. In this way, using (37) we obtain:

$$\sum_n \chi_{m,n} c_n = \mathbf{E}_{av} \cdot \int_\Omega \mathbf{w}_{m,-\mathbf{k}}(\mathbf{r}) e^{-j\mathbf{k}\cdot\mathbf{r}} d^3\mathbf{r} \tag{38a}$$

$$\chi_{m,n} = \int_\Omega \frac{1}{\varepsilon_r - 1} \mathbf{w}_{m,-\mathbf{k}}(\mathbf{r}) \cdot \mathbf{w}_{n,\mathbf{k}}(\mathbf{r}) d^3\mathbf{r} - \iint_{\Omega\Omega} \mathbf{w}_{m,-\mathbf{k}}(\mathbf{r}) \cdot \beta^2 \overline{\overline{\mathbf{G}}}_{p0}(\mathbf{r}|\mathbf{r}') \cdot \mathbf{w}_{n,\mathbf{k}}(\mathbf{r}') d^3\mathbf{r}\, d^3\mathbf{r}' \tag{38b}$$

Notice that the expansion functions vanish outside the dielectric inclusions, and so the integration domain may be replaced by $\{\mathbf{r} \in \Omega : \varepsilon_r(\mathbf{r}) - 1 \neq 0\}$. For convenience we denote the inverse of the matrix $[\chi_{m,n}]$ with generic element $\chi_{m,n}$ defined as above, as $[\chi^{m,n}]$. Substituting (37) in (16) and using (38) it is easy to verify that:

$$\frac{\hat{\mathbf{P}}(\mathbf{E})}{\varepsilon_0} = \left( \frac{1}{V_{cell}} \sum_{m,n} \chi^{m,n} \int_\Omega \mathbf{w}_{m,\mathbf{k}}(\mathbf{r}) e^{+j\mathbf{k}\cdot\mathbf{r}} d^3\mathbf{r} \int_\Omega \mathbf{w}_{n,-\mathbf{k}}(\mathbf{r}) e^{-j\mathbf{k}\cdot\mathbf{r}} d^3\mathbf{r} \right) \cdot \mathbf{E}_{av} \tag{39}$$

Therefore, using (12) we finally obtain the important result:

$$\frac{\overline{\overline{\varepsilon}}_{eff}}{\varepsilon_0}(\omega, \mathbf{k}) = \overline{\overline{\mathbf{I}}} + \frac{1}{V_{cell}} \sum_{m,n} \chi^{m,n} \int_\Omega \mathbf{w}_{m,\mathbf{k}}(\mathbf{r}) e^{+j\mathbf{k}\cdot\mathbf{r}} d^3\mathbf{r} \int_\Omega \mathbf{w}_{n,-\mathbf{k}}(\mathbf{r}) e^{-j\mathbf{k}\cdot\mathbf{r}} d^3\mathbf{r} \tag{40}$$

The formula is valid for dielectric crystals with no PEC surfaces. Equation (40) is quite interesting because it establishes a very simple relation between the dielectric function, the expansion functions, and the elements $\chi^{m,n}$. In the next section, we will illustrate the application of the formalism to some relevant microstructured materials with negative parameters.

In the particular case in which the material only contains PEC surfaces and $\varepsilon_r - 1 = 0$ in the rest of the unit cell, the unknown of the integral equation is taken equal to the vector tangential density $\mathbf{f} = \frac{1}{\beta^2} \hat{\mathbf{v}}' \times [\nabla' \times \mathbf{E}]$ defined over the metallic surface $\partial D$. The vector field $\mathbf{f}$ is proportional to the density of current $\mathbf{J}_c$ over the PEC surfaces. As in the dielectric case – see (37) – the unknown is expanded in terms of the complete set of vectors $\mathbf{w}_1$, $\mathbf{w}_2$, …, except that now the expansion functions form a complete set of tangential vector fields over the metallic surface. Proceeding as in the dielectric case and enforcing that the tangential component of the electric field vanishes at the PEC surface, it is possible to prove that the permittivity dyadic is:

$$\frac{\overline{\overline{\varepsilon_{eff}}}}{\varepsilon_0}(\omega,\mathbf{k}) = \overline{\overline{\mathbf{I}}} + \frac{1}{V_{cell}} \sum_{m,n} \chi^{m,n} \int_{\partial D} \mathbf{w}_{m,\mathbf{k}}(\mathbf{r}) e^{+j\mathbf{k}\cdot\mathbf{r}} ds \int_{\partial D} \mathbf{w}_{n,-\mathbf{k}}(\mathbf{r}) e^{-j\mathbf{k}\cdot\mathbf{r}} ds \qquad (41a)$$

$$\chi_{m,n} = \int_{\partial D}\int_{\partial D} \left( \nabla_s \cdot \mathbf{w}_{m,-\mathbf{k}}(\mathbf{r}) \nabla_s' \cdot \mathbf{w}_{n,\mathbf{k}}(\mathbf{r}') - \beta^2 \mathbf{w}_{m,-\mathbf{k}}(\mathbf{r}) \cdot \mathbf{w}_{n,\mathbf{k}}(\mathbf{r}') \right) \Phi_{p0}(\mathbf{r}|\mathbf{r}') \, ds \, ds' \qquad (41b)$$

In above $\nabla_s \cdot$ stands for the surface divergence of a tangential vector field, and, as before, the matrix $[\chi^{m,n}]$ is the inverse of $[\chi_{m,n}]$.

## V. CHARACTERIZATION OF METAMATERIALS WITH NEGATIVE PARAMETERS

In order to validate and illustrate the application the theoretical formalism developed in the previous sections, we will homogenize several configurations of artificial materials formed by split-ring resonators and metallic wires. Before that, it is worth mentioning that the homogenization approach that we used in [21] to characterize a three-dimensional array of connected and non-connected wires is formally equivalent to the homogenization theory presented in this paper, even though in [21] the concepts were presented in a completely different manner, and the theory was very specialized to the particular problem under study. Thus, it can be said that the results presented in [21] present further evidence that the homogenization theory developed here can be used to characterize the effective parameters of complex metamaterials.

Next, our objective is to characterize the effective permittivity and permeability of metamaterials formed by metallic wires and split ring resonators. As discussed in section II, this requires linking the dielectric function $\overline{\overline{\varepsilon_{eff}}}(\omega,\mathbf{k})$ – which can be computed numerically using (41) – with the parameters $\overline{\overline{\varepsilon_r}}(\omega)$, $\overline{\overline{\mu_r}}(\omega)$, etc, associated with the local model (3). As argued in section II, if such relation exists it is necessarily of the form (6). For simplicity, the examples analyzed here were chosen in such a way that the microstructure of the medium has enough symmetry so that the medium is non-gyrotropic, i.e. the first order derivatives of $\overline{\overline{\varepsilon_{eff}}}(\omega,\mathbf{k})$ with respect to $\mathbf{k}$ vanish at the origin, or equivalently the magneto-electric tensors $\overline{\overline{\xi}}$ and $\overline{\overline{\zeta}}$ vanish. To this end, instead of using the standard edge-side coupled split-ring resonator formed by two concentric rings, we will use the modified broadside coupled split-ring resonator formed by two parallel rings [23]. As proved in [23], this modified split-ring resonator (MSRR) does not permit magneto-electric coupling. The characterization of the bianisotropic properties of artificial media with relevant optical activity will be reported in a future communication.

Since, as discussed above, the magneto-electric tensors must vanish, we find from (6) that:

$$\overline{\overline{\varepsilon_r}}(\omega) = \lim_{k \to 0} \frac{\overline{\overline{\varepsilon_{eff}}}(\omega,\mathbf{k})}{\varepsilon_0} \qquad (42)$$

Moreover, in this work we assume that the split ring resonators are parallel to the *xoy* plane. This implies that the magnetic permeability of the artificial medium is of the form $\overline{\overline{\mu_r}}(\omega) = \hat{\mathbf{u}}_x\hat{\mathbf{u}}_x + \hat{\mathbf{u}}_y\hat{\mathbf{u}}_y + \mu_{zz}\hat{\mathbf{u}}_z\hat{\mathbf{u}}_z$. Substituting this formula into (6) and using (42), we easily find that,

$$\mu_{zz}(\omega) = \left(1 + \lim_{\substack{k_x \to 0 \\ k_y=k_z=0}} -\frac{\beta^2}{k_x^2}\hat{\mathbf{u}}_y \cdot \left(\frac{\overline{\overline{\varepsilon_{eff}}}}{\varepsilon_0}(\omega,\mathbf{k}) - \overline{\overline{\varepsilon_r}}(\omega)\right) \cdot \hat{\mathbf{u}}_y\right)^{-1} \qquad (43)$$

Equivalently, we can write that:

$$\mu_{zz}(\omega) = \frac{1}{1 - \beta^2 \frac{1}{2\varepsilon_0}\left.\frac{\partial^2 \varepsilon_{eff,yy}}{\partial k_x^2}\right|_{\mathbf{k}=0}} \qquad (44)$$

where $\varepsilon_{eff,yy} = \hat{\mathbf{u}}_y \cdot \overline{\overline{\varepsilon_{eff}}} \cdot \hat{\mathbf{u}}_y$. Thus, we found out that *provided the metamaterial can be described using a permittivity/permeability model* as in (3) then its constitutive parameters are given by (42) and (44). Notice that consistently with the observations of section II and with the results of [17, Sect. 103], the magnetic permeability is a function of the second derivatives of the dielectric function with respect to the wave vector.

In the first example, we characterize the effective parameters of a metamaterial similar to the one proposed in the seminal work of Smith et al [2], formed by MSRRs and metallic wires (see the inset of Fig. 3). The distance between the inclusions along the coordinate axes is $a_x = a_y \equiv a$, and $a_z = 0.5a$ (note that the lattice is not simple cubic). As referred before, the MSRRs are parallel to the *xoy*-plane, and are formed by two rings with mean radius $R_{med} = 0.4a$ with an angular gap of 10[deg]. To simplify the numerical implementation of the proposed homogenization method, we assumed that the rings are formed by thin metallic wires with circular cross-section and radius $0.01a$, instead of being planar particles as proposed in [23]. The distance between the two rings (relatively to the mid-plane of each ring) is $d = 0.125a$. On the other hand, the continuous metallic wires are direct along the *y*-direction (see the inset of Fig. 3) and also have radius $0.01a$.

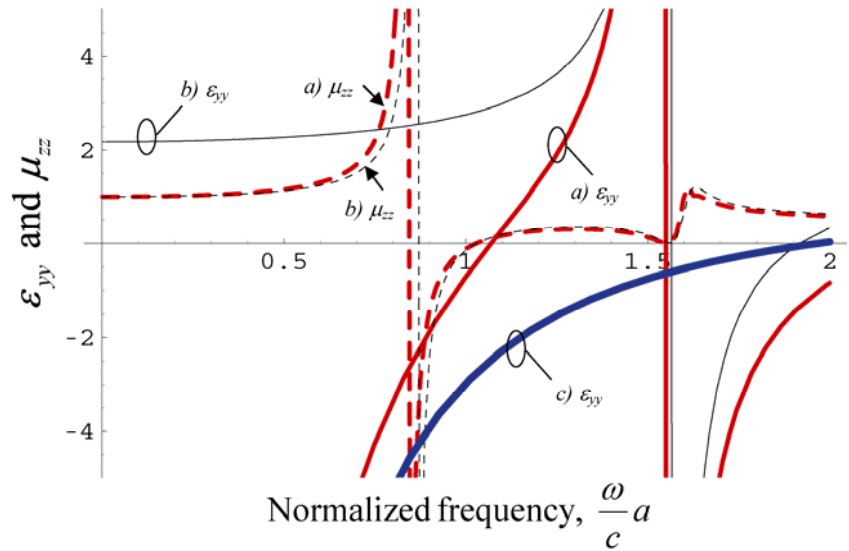

**Fig. 2** (Color online) Extracted effective permittivity (solid line) and effective permeability (dashed line) for a metamaterial formed by *a*) continuous wires + MSRRs (medium thick red lines – light gray in grayscale), *b*) only MSRRs (thin black lines), and *c*) only continuous wires (thick blue line – dark gray in grayscale).

Using (41), (42) and (44) we have computed the constitutive parameters $\bar{\bar{\varepsilon}}_r(\omega)$ and $\mu_{zz}$ of the composite medium. To this end, we have expanded the density of electric current $\mathbf{J}_c$ in an appropriate set of expansion functions $\mathbf{w}_n = \mathbf{w}_{n,\mathbf{k}}(\mathbf{r})$ tangential to the PEC surfaces, and then we used (41) to calculate the dielectric function of the composite medium. For simplicity, to ease the numerical calculation of the matrix $[\chi_{m,n}]$, we have used the well-known thin-wire approximation [31]-[32], which assumes that the current along each wire flows exclusively along its axis. Typically, we used 4-5 expansion functions per wire/ring to obtain the numerical results. The numerical derivatives that appear in (44) where evaluated using numerical methods.

The extracted effective permittivity $\varepsilon_{yy}$ and effective permeability $\mu_{zz}$ are depicted in Fig. 2 for three configurations of the metamaterial. The permittivity along $z$, $\varepsilon_{zz} = 1$, is not depicted in the figure. Note that the extracted parameters are real numbers because we assume that the wires are perfectly conducting. Consistently with the results of [2], it is found that the effective permittivity is negative when the MSRRs are removed and the material is formed by continuous wires (curve *c*). On the other hand, if the continuous wires are removed and the metamaterial is formed by only MSRRs (curve *b*), the numerical results predict that the permeability has a resonance at the normalized frequency $\omega a / c = 0.86$. When the MSRRs and the continuous wires are combined (curve *a*) there is a frequency window, $0.86 < \omega a / c < 1.04$, where both the effective permittivity and permeability are simultaneously negative. In order to confirm these results and check the accuracy of the proposed homogenization method, we have calculated the band structure of the composite medium using the extracted $\varepsilon_{yy}$ and $\mu_{zz}$. For propagation along the *x*-direction, we should have the dispersion relation $k_x = \pm \frac{\omega}{c} \sqrt{\varepsilon_{yy}(\omega) \mu_{zz}(\omega)}$. Solving this equation with respect to

$\omega$, we obtain several bands of the form $\omega = \omega(k_x)$, which are depicted in Fig. 3 (solid black lines). Then, we compared these results (based on the proposed homogenization method) with the "exact" band structure calculated using the full wave hybrid method introduced in [32] ("star symbols" in Fig. 3). The full wave results can be considered "numerically exact", even though we should refer that we also used a thin-wire approximation to simplify the implementation the full-wave method proposed [32].

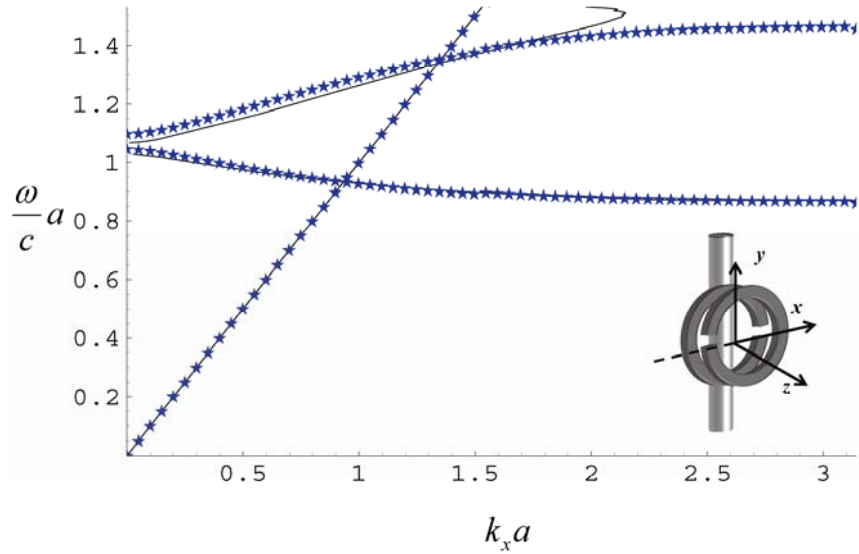

**Fig. 3** (Color online) Band structure of composite material formed by wires + MSRRs (geometry is shown in the inset). The solid black line was calculated using the extracted $\varepsilon_{yy}$ and $\mu_{zz}$. The "star"-symbols were obtained using the full wave hybrid method introduced in [32].

It is seen in Fig. 3 that the comparison between the homogenization results and the full wave results is good, particularly for frequencies such that $\omega a/c < 1.3$. In particular, the frequency band where the material has both permittivity and permeability simultaneously negative is predicted with very good accuracy, even for values of $k_x$ near the edge of the Brillouin zone. This is a very interesting and unexpected result, because in general the scope of application of homogenization methods is thought to be $\omega a/c \ll \pi$ and $ka/c \ll \pi$. For frequencies above $\omega a/c = 1.4$, near the resonance of $\varepsilon_{yy}$, the agreement quickly deteriorates and the magnetic permeability ceases to have meaning (notice that when $\varepsilon_{yy}$ varies fast it is not expected that a

Taylor expansion of the dielectric function can be accurate, and thus spatial dispersion becomes dominant).

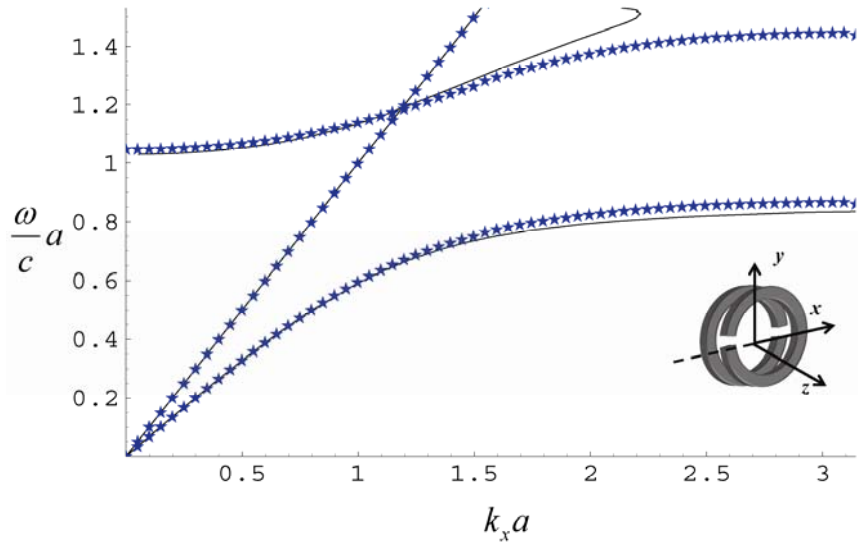

**Fig. 4** (Color online) Band structure of composite material formed by MSRRs (geometry is shown in the inset). The solid black line was calculated using the extracted $\varepsilon_{yy}$ and $\mu_{zz}$. The "star"-symbols were obtained using the full wave hybrid method introduced in [32].

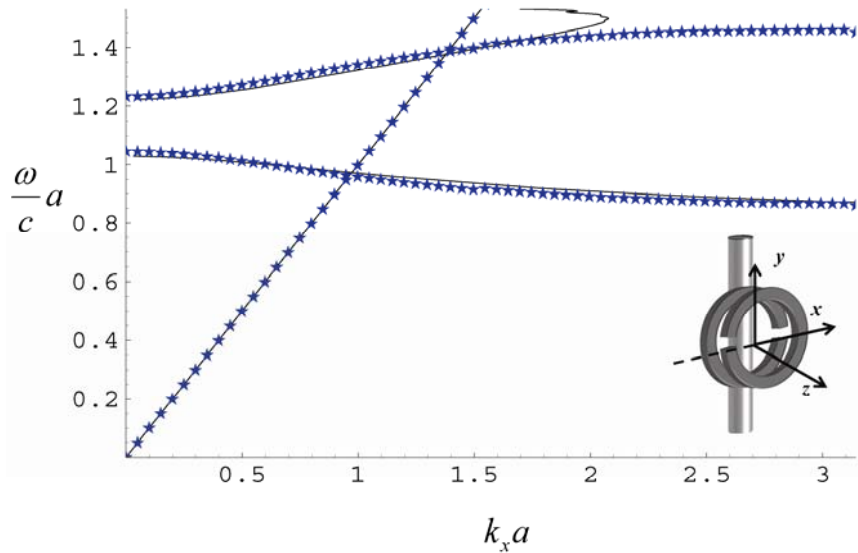

**Fig. 5** (Color online) Same as Fig. 3 but the radius of the continuous wires (directed along *y*) is increased 5 times.

We have performed similar band structure calculations for the case in which the continuous wires are removed, and the material is formed uniquely by MSRRs. These results are reported in Fig. 4. Notice that as in Smith's original work [2], the frequency region where the composite material has simultaneously negative parameters, becomes a frequency band gap when the metallic wires are removed.

The effective permittivity of the composite medium can be tuned by varying the radius of the continuous wires. For example, if one increases the radius 5 times (radius $0.05a$) the band structure results obtained with the extracted effective parameters (not shown here for the sake of brevity) and the full wave numerical method [32], are depicted in Fig. 5. Comparing with Fig. 3, it is seen that the gap between the band where the material has both $\varepsilon_{yy}$ and $\mu_{zz}$ negative, and the band where $\varepsilon_{yy}$ and $\mu_{zz}$ are simultaneously positive, becomes more pronounced, confirming the fact that $\varepsilon_{yy}$ becomes more negative when the radius of the wires increases.

We have also investigated the magnetic properties of media formed by elliptical-rings. The motivation is that we intuitively expect that by using rings with large axial ratios, so that the semi-axis of the ring along the *y*-direction $R_y$ is significantly larger than the semi-axis $R_x$ along the *x*-direction, it may be possible to decrease the electrical size (footprint in the *xoz* plane) of the particles. This may have interesting applications in problems where the propagation is limited to the *xoz* plane. Notice that the resolution of sub-wavelength imaging devices based on metamaterials is limited by the electrical size of the particles, and so there is a great interest in designing more compact resonant particles and thus improve the performance of the devices.

In our simulations we assumed that the distance between the inclusions along the coordinate axes is $a_x \equiv a$, $a_y = 2a$, and $a_z = 0.5a$. The elliptical MSRRs are parallel to the *xoy*-plane, and are formed by two elliptical rings (see the inset of Fig. 7). The (mean) semi-axes of the rings are $R_x = 0.4a$ and $R_y = 0.8a$ (the axial ratio is 2). The angular gap is

42.5[deg]. As before, the distance between the rings is $d = 0.125a$ and all the metallic wires have radius $0.01a$.

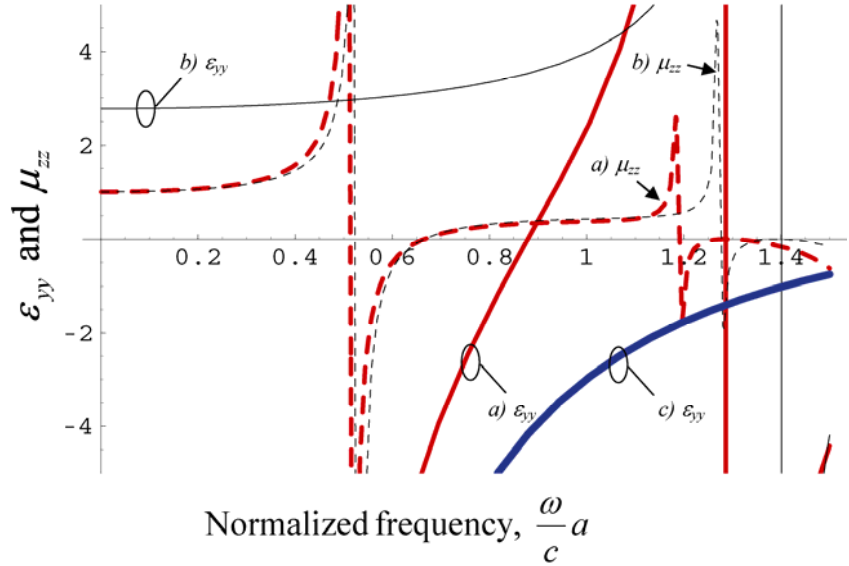

**Fig. 6** (Color online) Extracted effective permittivity (solid line) and effective permeability (dashed line) for a metamaterial formed by *a*) continuous wires + elliptical MSRRs (medium thick red lines – light gray in grayscale), *b*) only elliptical MSRRs (thin black lines), and *c*) only continuous wires (thick blue line – dark gray in grayscale).

In Fig. 6 the extracted effective parameters are depicted for composite media with both elliptical rings and wires (curve *a*), only elliptical rings (curve *b*), and only wires (curve *c*). The results are qualitatively similar to those reported for circular rings. There is however an important novelty: the resonance of the magnetic particles is now at $\omega a/c = 0.5$, i.e. the resonant frequency is only 58% of that we obtained for circular rings. Moreover, the frequency band where the composite medium has both $\varepsilon_{yy}$ and $\mu_{zz}$ negative is now $0.5 < \omega a/c < 0.65$ and so the absolute bandwidth associated with the negative refraction effect is nearly the same as that for circular rings, whereas the relative bandwidth (the most important parameter) is almost twice. These results confirm our intuition and demonstrate that elliptically shaped resonant rings may have interesting potentials and may help reducing the electrical size of the inclusions. The comparison between the band structure computed using the extracted $\varepsilon_{yy}$ and $\mu_{zz}$ parameters and the full wave method proposed in [32], is shown in Fig. 7 and Fig. 8, revealing a very good agreement for $\omega a/c < 1.1$, and confirming that for low frequencies the metamaterial can be described using a local homogenization model.

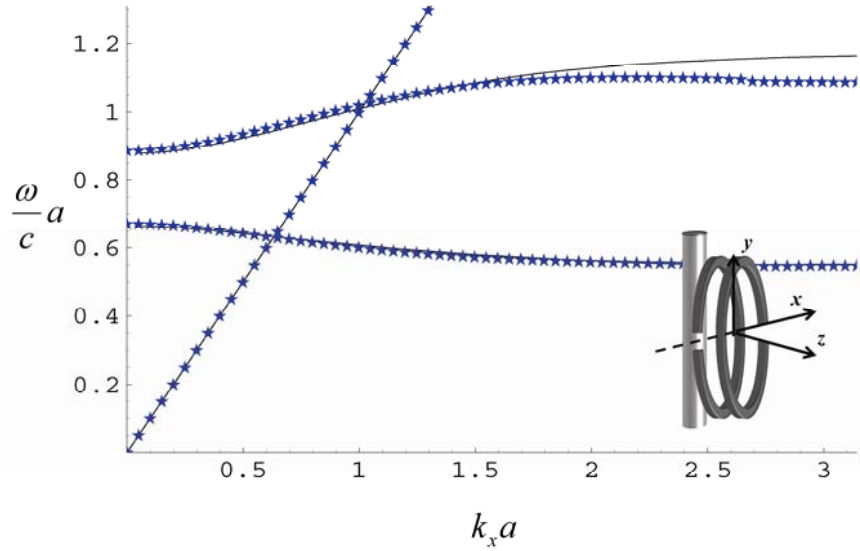

**Fig. 7** (Color online) Band structure of composite material formed by wires + elliptical MSRRs (geometry is shown in the inset). The solid black line was calculated using the extracted $\varepsilon_{yy}$ and $\mu_{zz}$. The "star"-symbols were obtained using the full wave hybrid method introduced in [32].

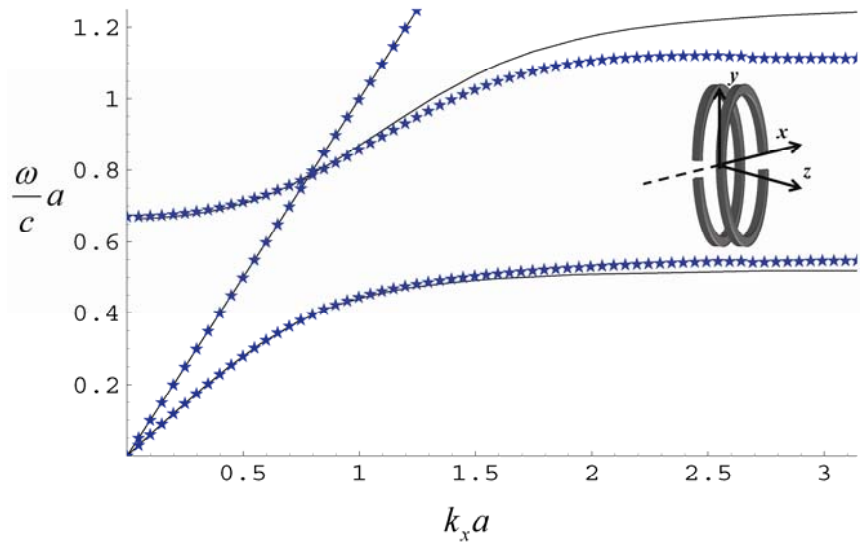

**Fig. 8** (Color online) Band structure of composite material formed by elliptical MSRRs (geometry is shown in the inset). The solid black line was calculated using the extracted $\varepsilon_{yy}$ and $\mu_{zz}$. The "star"-symbols were obtained using the full wave hybrid method introduced in [32].

In the last example, we investigate a different opportunity to reduce the electrical size of the magnetic particles. In [33] it was shown that the bandwidth of the negative refraction regime in a material formed by plasmonic spheres may be improved by closely-packing the inclusions, and in this way enhancing the mutual coupling. Extrapolating this result to our problem, we may expect that if the MSRR rings are closely packed (along the *z*-direction; see the inset of Fig. 9), it may be possible to enhance artificial magnetism, and in this way reduce the electrical size of the particles. A similar configuration has also been studied in [34], where it was found that the permeability of the metamaterial could be reasonably high over a wide frequency range. Also in [35], planar near-field magnetoinductive lens based on such geometry was investigated. It was proved that due to the excitation of magnetoinductive surface waves [36] it was possible to achieve sub-wavelength imaging. In a certain sense the structure studied in [35] can be regarded as the magnetic dual of the wire medium [37]. Here, we will characterize the effective parameters of such metamaterial as an application of the proposed homogenization method.

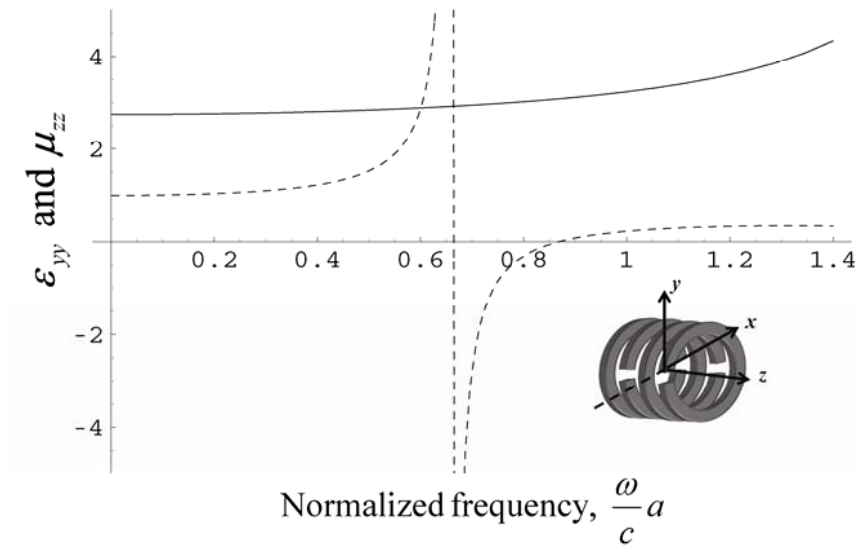

**Fig. 9** Extracted effective permittivity (solid line) and effective permeability (dashed line) for a metamaterial formed by closely packed MSRRs (geometry is shown in the inset).

As in the first example, we suppose that the rings have a circular geometry. The dimensions of rings are the same in the first example, but now the lattice constants are $a_x = a_y \equiv a$, and $a_z = 0.25a$, i.e. as compared to the first example the distance between to MSRR rings along $z$ is reduced to half, increasing in this way the capacitance between the rings. The extracted effective permittivity (solid line) and permeability (dashed line) are shown in Fig. 9. The material exhibits negative permeability in the frequency range $0.66 < \omega a/c < 0.87$, i.e. the resonant frequency is about 77% of that obtained in the first example while the absolute bandwidth increases 16%. This confirms that by reducing the distance between adjacent MSRRs, it may indeed be possible to design more compact metamaterials with electrically smaller particles. The comparison between the band structures calculated using the extracted parameters and the full wave method [32] is depicted in Fig. 10, further supporting and validating the proposed homogenization theory.

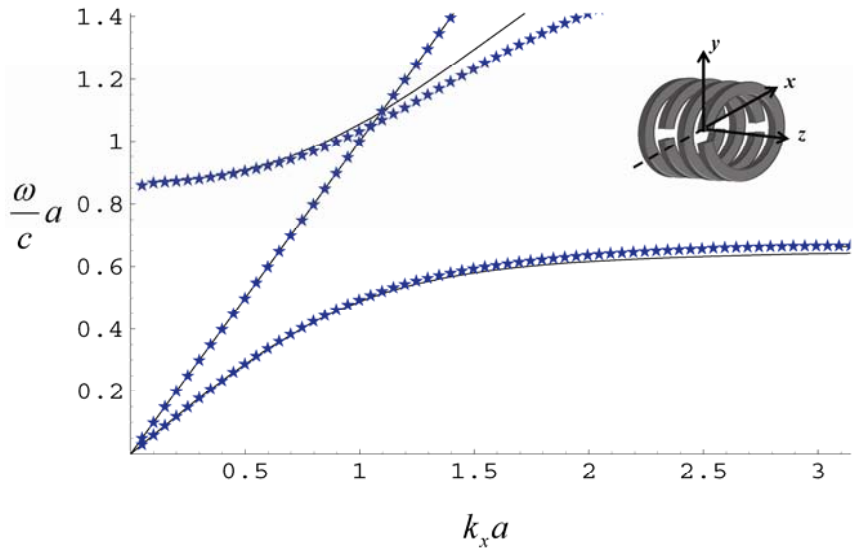

**Fig. 10** (Color online) Band structure of composite material formed by closely packed MSRRs (geometry is shown in the inset). The solid black line was calculated using the extracted $\varepsilon_{yy}$ and $\mu_{zz}$. The "star"-symbols were obtained using the full wave hybrid method introduced in [32].

# VI. CONCLUSION

In this work we developed a numerical method to homogenize arbitrary non-magnetic periodic materials. In our formalism the homogenized medium is described by a dielectric function of the form $\overline{\overline{\varepsilon}}_{eff} = \overline{\overline{\varepsilon}}_{eff}(\omega, \mathbf{k})$. The advantage of this approach is that it is completely general and allows characterizing both frequency and spatial dispersion. It was discussed that if a medium is local, i.e. if it can be characterized using effective parameters $\overline{\overline{\varepsilon}}_r(\omega)$ and $\overline{\overline{\mu}}_r(\omega)$ (and possibly magneto-electric crossed terms), it is possible to link these local parameters with the dielectric function $\overline{\overline{\varepsilon}}_{eff}(\omega, \mathbf{k})$. Namely, it was discussed that the effective permeability may be related with the second order derivatives of the dielectric function with respect to the wave vector, consistently with [17, Sect. 103].

It was proved that the problem of calculation of the dielectric function $\overline{\overline{\varepsilon}}_{eff}(\omega, \mathbf{k})$ can be reduced to an integral-differential system. We developed an approach based on the MoM to obtain the numerical solution of such system. In particular, we found out that the dielectric function can be written in closed analytical form – see formulas (40) and (41) – in terms of the inverse of an infinite matrix. In order to validate the proposed concepts, we studied numerically the properties of media formed by metallic wires and split-ring resonators. It was explained how to extract $\overline{\overline{\varepsilon}}_r(\omega)$ and $\overline{\overline{\mu}}_r(\omega)$ from the non-local dielectric function. The extracted effective permittivity and permeability were used to predict the dispersion characteristics $\omega = \omega(\mathbf{k})$, showing good agreement with results obtained using a full wave method that calculates the band structure of periodic media. In order to investigate the possibility of designing more compact metamaterials, we characterized microstructured composites formed by elliptical split-rings and closely packed rings. The results of the simulations showed good potentials for these structures, and that the size of the inclusions may be significantly reduced. This may have important applications in subwavelength imaging devices. To conclude, we underline that the proposed homogenization method can be applied to a wide range of composite structures, and can also be used to characterize bianisotropic media (e.g. chiral media). These results will be reported in a future work [38], where the extracted parameters are applied to study a reflectivity problem.

## APPENDIX A

In this Appendix, the results enunciated in section III.D are proved. To begin with, we prove property P1). To this end, let $\mathbf{E}_1$ be a solution of the regularized homogenization problem (20) associated with $(\mathbf{k}, \omega, \mathbf{E}_{av,1})$ and $\mathbf{E}_2$ be a solution associated with $(-\mathbf{k}, \omega, \mathbf{E}_{av,2})$. Then, using (20) we have that:

$$\nabla \cdot \left( -\mathbf{E}_1 \times \nabla \times \mathbf{E}_2 + \mathbf{E}_2 \times \nabla \times \mathbf{E}_1 \right) =$$
$$\beta^2 \frac{1}{\varepsilon_0} \Big[ \mathbf{E}_1 e^{j\mathbf{k}\cdot\mathbf{r}} \cdot \left( \hat{\mathbf{P}}_{av}\left(\mathbf{E}_{av,2};-\mathbf{k}\right) - \hat{\mathbf{P}}\left(\mathbf{E}_2;-\mathbf{k}\right) \right) \qquad (A1)$$
$$- \mathbf{E}_2 e^{-j\mathbf{k}\cdot\mathbf{r}} \cdot \left( \hat{\mathbf{P}}_{av}\left(\mathbf{E}_{av,1};\mathbf{k}\right) - \hat{\mathbf{P}}\left(\mathbf{E}_1;\mathbf{k}\right) \right) \Big]$$

Next, we integrate both sides of the equation over the unit cell. Using the divergence theorem, the Floquet boundary conditions, and the boundary conditions at the dielectric/metallic interfaces, it is clear that the integral corresponding to the left-hand side term vanishes. Therefore, using (12) and the fact that the spatial average of $\mathbf{E}_i$ is $\mathbf{E}_{av,i}$, $i=1,2$, we find that,

$$\mathbf{E}_{av,1} \cdot \left( \frac{\hat{\mathbf{P}}_{av}}{\varepsilon_0}\left(\mathbf{E}_{av,2};-\mathbf{k}\right) - \left( \frac{\overline{\overline{\varepsilon_{eff}}}}{\varepsilon_0}(\omega,-\mathbf{k}) - \overline{\overline{\mathbf{I}}} \right) \cdot \mathbf{E}_{av,2} \right) = \mathbf{E}_{av,2} \cdot \left( \frac{\hat{\mathbf{P}}_{av}}{\varepsilon_0}\left(\mathbf{E}_{av,1};\mathbf{k}\right) - \left( \frac{\overline{\overline{\varepsilon_{eff}}}}{\varepsilon_0}(\omega,\mathbf{k}) - \overline{\overline{\mathbf{I}}} \right) \cdot \mathbf{E}_{av,1} \right)$$
(A2)

Using (17) and noting that the above equation holds for arbitrary $\mathbf{E}_{av,1}$ and $\mathbf{E}_{av,2}$, we easily obtain P1).

Next, we demonstrate property P2). The transformed metamaterial is characterized by the relative permittivity $\varepsilon'_r(\mathbf{r}') = \varepsilon_r(\mathbf{r})$, where $\mathbf{r}' = \mathbf{T}\cdot\mathbf{r}$, $\mathbf{T}:\mathbf{r} \to \mathbf{r} + \mathbf{u}$ is a translation along the constant vector $\mathbf{u}$, and $\varepsilon_r$ is the relative permittivity of the original periodic material. The metallic surfaces are transformed similarly. We define $\mathbf{E}'(\mathbf{r}') = \mathbf{E}(\mathbf{r})$ and $\mathbf{B}'(\mathbf{r}') = \mathbf{B}(\mathbf{r})$. Then, from (20) and using the identity $\nabla = \nabla'$, we find that:

$$\nabla' \times \mathbf{E}' = -j\omega \mathbf{B}' \qquad (A3a)$$

$$\nabla' \times \frac{\mathbf{B}'}{\mu_0} = j\omega \left( \hat{\mathbf{P}}_{av}(\mathbf{E}_{av}) - \hat{\mathbf{P}}(\mathbf{E}) \right) e^{j\mathbf{k}.\mathbf{u}} e^{-j\mathbf{k}.\mathbf{r}'} + \varepsilon_0 \varepsilon_r' j\omega \mathbf{E}' \qquad (A3b)$$

Because the volume of the transformed unit cell is the same as the original volume, it is clear that $\hat{\mathbf{P}}_{av} = \hat{\mathbf{P}}'_{av}$. On the other hand, using (16) and making a coordinate transformation in the integrals, we obtain that:

$$\frac{\hat{\mathbf{P}}(\mathbf{E})}{\varepsilon_0} = \frac{e^{-j\mathbf{k}.\mathbf{u}}}{V_{cell}} \left( \frac{1}{\beta^2} \int_{\mathbf{T}.\partial D} \hat{\mathbf{v}}' \times [\nabla' \times \mathbf{E}'] e^{+j\mathbf{k}.\mathbf{r}'} ds' + \int_{\mathbf{T}.\Omega} (\varepsilon_r' - 1) \mathbf{E}' e^{+j\mathbf{k}.\mathbf{r}'} d^3\mathbf{r}' \right) \qquad (A4)$$

In the above we used the fact that $\hat{\mathbf{v}}(\mathbf{r}) = \hat{\mathbf{v}}'(\mathbf{r}')$. Since the sets $\mathbf{T}.\partial D$ and $\mathbf{T}.\Omega$ are equivalent to $\partial D'$ and $\Omega'$, respectively, and because the integrands are periodic functions, we find that:

$$\hat{\mathbf{P}}(\mathbf{E}) = \hat{\mathbf{P}}'(\mathbf{E}') e^{-j\mathbf{k}.\mathbf{u}} \qquad (A5)$$

Substituting the above expression into (A3), it follows that $\mathbf{E}'$ is a solution of the homogenization problem in the transformed periodic material, associated with $(\mathbf{k}, \omega, \mathbf{E}_{av} e^{j\mathbf{k}.\mathbf{u}})$. Then, using (12) and (A5) it is immediate that P2) holds.

Now we consider that the transformed metamaterial is the result of the application of an isometry $\mathbf{S}$ (composition of rotations and reflections) to the original structure. The relative permittivity of the transformed structure is such that $\varepsilon_r'(\mathbf{r}') = \varepsilon_r(\mathbf{r})$, where $\mathbf{r}' = \mathbf{S}.\mathbf{r}$, and $\mathbf{S}$ is the isometry. We define $\mathbf{E}'(\mathbf{r}') = \mathbf{S}.\mathbf{E}(\mathbf{r})$ and $\mathbf{B}'(\mathbf{r}') = \det(\mathbf{S})\mathbf{S}.\mathbf{B}(\mathbf{r})$, where $\det(\mathbf{S})$ is the determinant of the application. Note that $\det(\mathbf{S})^2 = 1$ because $\mathbf{S}.\mathbf{S}^t = \bar{\bar{\mathbf{I}}}$. To proceed we need the following auxiliary formula (which is valid because $\mathbf{S}$ is an isometry):

$$\mathbf{S}.\nabla \times = \det(\mathbf{S})\nabla' \times \mathbf{S}. \qquad (A6)$$

Using this result and applying $\mathbf{S}$ to both sides of (20) we obtain:

$$\nabla \times \mathbf{E}' = -j\omega \mathbf{B}' \qquad (A7a)$$

$$\nabla \times \frac{\mathbf{B}'}{\mu_0} = j\omega \left( \mathbf{S}.\hat{\mathbf{P}}_{av}(\mathbf{E}_{av};\mathbf{k}) - \mathbf{S}.\hat{\mathbf{P}}(\mathbf{E};\mathbf{k}) \right) e^{-j(\mathbf{S}.\mathbf{k}).\mathbf{r}'} + \varepsilon_0 \varepsilon_r' j\omega \mathbf{E}' \qquad (A7b)$$

The volume of the transformed unit cell is invariant because $\mathbf{S}$ is an isometry. It is also straightforward to verify that $\mathbf{S}.\hat{\mathbf{P}}_{av}(\mathbf{E}_{av};\mathbf{k}) = \hat{\mathbf{P}}'_{av}(\mathbf{S}.\mathbf{E}_{av};\mathbf{S}.\mathbf{k})$. On the other hand, from (16) we obtain:

$$\frac{\mathbf{S}.\hat{\mathbf{P}}(\mathbf{E};\mathbf{k})}{\varepsilon_0} = \frac{1}{V_{cell}} \left( \frac{1}{\beta^2} \int_{\partial D} \mathbf{S}.\left( \hat{\mathbf{v}} \times [\nabla \times \mathbf{E}] \right) e^{+j(\mathbf{S}.\mathbf{k}).\mathbf{r}'} ds + \int_{\Omega} (\varepsilon_r' - 1) \mathbf{E}' e^{+j(\mathbf{S}.\mathbf{k}).\mathbf{r}'} d^3\mathbf{r} \right) \qquad (A8)$$

But, because $\mathbf{S}$ is an isometry, we have $\mathbf{S}.\left( \hat{\mathbf{v}} \times [\nabla \times \mathbf{E}] \right) = \det(\mathbf{S})\left( \mathbf{S}.\hat{\mathbf{v}} \times [\mathbf{S}.\nabla \times \mathbf{E}] \right)$. Using also (A6) and the fact that $\mathbf{S}.\hat{\mathbf{v}}(\mathbf{r}) = \hat{\mathbf{v}}'(\mathbf{r}')$, it is clear that $\mathbf{S}.\left( \hat{\mathbf{v}} \times [\nabla \times \mathbf{E}] \right) = \hat{\mathbf{v}}' \times [\nabla' \times \mathbf{E}']$. Substituting this result into (A8) and making a coordinate transformation in the integration variables, we obtain:

$$\frac{\mathbf{S}.\hat{\mathbf{P}}(\mathbf{E};\mathbf{k})}{\varepsilon_0} = \frac{1}{V_{cell}} \left( \frac{1}{\beta^2} \int_{\mathbf{S}.\partial D} \hat{\mathbf{v}}' \times [\nabla' \times \mathbf{E}'] e^{+j(\mathbf{S}.\mathbf{k}).\mathbf{r}'} ds' + \int_{\mathbf{S}.\Omega} (\varepsilon_r' - 1) \mathbf{E}' e^{+j(\mathbf{S}.\mathbf{k}).\mathbf{r}'} d^3\mathbf{r}' \right) \qquad (A9)$$

As before, the sets $\mathbf{S}.\partial D$ and $\mathbf{S}.\Omega$ are equivalent to $\partial D'$ and $\Omega'$, respectively. Therefore, since the integrands are periodic functions, it follows that:

$$\mathbf{S}.\hat{\mathbf{P}}(\mathbf{E};\mathbf{k}) = \hat{\mathbf{P}}'(\mathbf{E}';\mathbf{S}.\mathbf{k}) \qquad (A10)$$

Substituting the above expression into (A7), we conclude that $\mathbf{E}'$ is a solution of the homogenization problem in the transformed periodic material, associated with $(\mathbf{S}.\mathbf{k}, \omega, \mathbf{S}.\mathbf{E}_{av})$. Finally, using (12) and (A10) it is immediate that P3) holds.

To conclude, we note that P4) is an immediate consequence of P2) and P3).

## ACKNOWLEDGEMENT

This work was funded by Fundação para Ciência e a Tecnologia.